\documentclass[sigconf]{acmart}
\usepackage{algorithm}
\usepackage{algpseudocode}
\usepackage{algorithmicx}
\usepackage{graphicx}
\usepackage{amsmath, amsfonts}
\usepackage[toc,page]{appendix}
\usepackage{array}
\usepackage{booktabs}
\usepackage{color}
\usepackage{enumitem}
\usepackage[utf8]{inputenc} 
\usepackage{listings}
\usepackage{longtable}
\usepackage{mdframed}
\usepackage{multirow}
\usepackage[final]{pdfpages}
\usepackage{pgfplots}	
\pgfplotsset{compat = 1.14}
\usepackage{rotating}
\usepackage{tabu}
\usepackage{tabularx}
\usepackage{tabulary}
\usepackage{textcomp}
\usepackage{tikz}
\usetikzlibrary{matrix,shapes,arrows,positioning,chains, calc, automata,arrows.meta}
\usepackage{todonotes} % add option disable to remove todos in document
\usepackage{stmaryrd}
\usepackage{rotating}
\usepackage{pdflscape}

\newcommand\copyrightnotice[1]{
	\begin{tikzpicture}[remember picture,overlay]
	\node[anchor=south,yshift=10pt] at (current page.south) {\fbox{\parbox{\dimexpr\textwidth-\fboxsep-\fboxrule\relax}{#1}}};
	\end{tikzpicture}
}


\makeatletter
\newcounter{protocol}%[chapter] 
\newenvironment{protocol}[1][]{
	\let\c@algorithm\c@protocol
	\makeatletter
	\renewcommand{\ALG@name}{Protocol}
	\makeatother
	\begin{algorithm}[#1]
	}{\end{algorithm}
}
\makeatother

\makeatletter
\newcounter{protocolDouble}%[chapter] 

\makeatother

\makeatletter
\newcounter{gate}%[chapter] 
\newenvironment{gate}[1][]{
	
	\let\c@algorithm\c@gate
	\renewcommand{\ALG@name}{Gate}
	
	\begin{algorithm}[#1]
	}{\end{algorithm}
}
\makeatother

\makeatletter
\newcounter{alg}%[chapter] 

\setlist[enumerate]{label* = \arabic*.}
\setlistdepth{9}
\newlist{myEnumerate}{enumerate}{9}
\setlist[myEnumerate]{label* = \arabic*.}

\newcolumntype{L}[1]{>{\raggedright\let\newline\\\arraybackslash\hspace{0pt}}m{#1}}
\newcolumntype{C}[1]{>{\centering\let\newline\\\arraybackslash\hspace{0pt}}m{#1}}
\newcolumntype{R}[1]{>{\raggedleft\let\newline\\\arraybackslash\hspace{0pt}}m{#1}}

\tikzset{every tree node/.style={align=center}}

\DeclareMathAlphabet{\pazocal}{OMS}{zplm}{m}{n}

\newcommand*{\patient}[1]{\ensuremath{\textbf{p}^{(#1)}}}
\newcommand*{\donor}[1]{\ensuremath{\textbf{d}^{(#1)}}}
\newcommand*{\donorbloodvec}[1]{\ensuremath{B^{d}_{#1}}}
\newcommand*{\patientbloodvec}[1]{\ensuremath{B^{p}_{#1}}}
\newcommand*{\antigenvec}[1]{\ensuremath{A^{d}_{#1}}}
\newcommand*{\antibodyvec}[1]{\ensuremath{A^{p}_{#1}}}
\newcommand*{\kequote}[1]{\ensuremath{\mathbf{q}^{(#1)}}}
\newcommand*{\ecg}{\ensuremath{G^{\textnormal{EC}}}}
\newcommand*{\ecsg}{\ensuremath{\mathbb{G}^{\textnormal{EC}}}}
\newcommand*{\pecg}{\ensuremath{G^{\textnormal{PEC}}}}
\newcommand*{\pecsg}{\ensuremath{\mathbb{G}^{\textnormal{PEC}}}}
\newcommand*{\aecg}{\ensuremath{G^{\textnormal{SEC}}}}
\newcommand*{\out}{\ensuremath{o}}

\definecolor{smartgray}{RGB}{61, 62, 64}
\definecolor{smartblue}{HTML}{0B6296}
\definecolor{smartred}{HTML}{C54949}
\definecolor{smartyellow}{HTML}{d0d62a}
\definecolor{smartgreen}{HTML}{61c15d}
\definecolor{smartorange}{HTML}{ff7e00}
\definecolor{smartviolet}{HTML}{d19fe8}
\definecolor{smartbrown}{HTML}{480607}
\definecolor{smartcadet}{HTML}{5F9EA0}

\newcommand*{\ie}{i.e.}

\newcommand*{\set}{\ensuremath{:=}}

\newcommand*{\primeset}{\ensuremath{\mathbf{P}}}

\newcommand*{\iverson}[1]{\ensuremath{[#1]}}

\newcommand*{\interval}[1]{\ensuremath{I^{(#1)}}}

\newcommand*{\nats}{\ensuremath{\mathbb{N}}}
\newcommand*{\natz}{\ensuremath{\mathbb{N}_{0}}}
\newcommand*{\natsbnd}[1]{\ensuremath{\mathbb{N}^{#1}}}

\newcommand*{\draw}{\ensuremath{\leftarrow}}
\newcommand*{\rdraw}{\ensuremath{\leftarrow_{\$}}}
\newcommand*{\rset}{\ensuremath{\leftarrow_{\$}}}

\newcommand*{\primeNumber}{\ensuremath{p}}

\newcommand*{\bigo}[1]{\ensuremath{\mathcal{O}(#1)}}

\newcommand*{\enc}[1]{\ensuremath{\llbracket #1 \rrbracket}}

\newcommand*{\dece}{\ensuremath{\textnormal{\textsf{Dec}}}}

\newcommand*{\plainspace}{\ensuremath{\mathbb{P}}}
\newcommand*{\cipherspace}{\ensuremath{\mathbb{C}}}
\newcommand*{\addh}{\ensuremath{+_h}}

\newcommand*{\subh}{\ensuremath{-_h}}

\newcommand*{\crossh}{\ensuremath{\times_h}}

\newcommand*{\blind}[1]{\ensuremath{\textnormal{\textsf{Blind}}(#1)}}

\newcommand*{\partysymbol}[1]{\ensuremath{P_{#1}}}
\newcommand*{\parties}{\ensuremath{\mathcal{P}}}

\newcommand*{\numparties}{\ensuremath{\iota}}
\newcommand*{\partythresh}{\ensuremath{\tau}}
\newcommand*{\numcorparties}{\ensuremath{\kappa}}
\newcommand*{\corparties}{\ensuremath{C}}

\newcommand*{\emptystr}{\ensuremath{\bot}}
\newcommand*{\prot}{\ensuremath{\pi}}
\newcommand*{\func}{\ensuremath{\pazocal{F}}}
\newcommand*{\gat}{\ensuremath{\rho}}
\newcommand*{\gatfunc}{\ensuremath{\pazocal{G}}}
\newcommand*{\simu}{\ensuremath{\textnormal{\textsf{S}}}}

\newcommand*{\secparam}{\ensuremath{s}}
\newcommand*{\simval}[1]{\ensuremath{\langle #1 \rangle}}

\newcommand*{\qu}[1]{\ensuremath{\mathbf{q}^{(#1)}}}

\newcommand*{\cyclesize}{\ensuremath{m}}

\newcommand*{\degree}[1]{\ensuremath{deg(#1)}}
\newcommand*{\indeg}[1]{\ensuremath{deg^+(#1)}}
\newcommand*{\outdeg}[1]{\ensuremath{deg^-(#1)}}
\newcommand*{\cg}{\ensuremath{G^{\text{C}}}}

\newcommand*{\welfare}{\ensuremath{\pazocal{W}}}

\newcommand*{\ndonor}[1]{\ensuremath{N^{(#1)}_{d}}}

\newcommand*{\npatient}[1]{\ensuremath{N^{(#1)}_{p}}}

\newcommand*{\ndonorpatient}[1]{\ensuremath{N^{(#1)}_{d,p}}}
\newcommand*{\ndonorpatientstar}[1]{\ensuremath{N^{*(#1)}_{d,p}}}
\newcommand*{\ndonorpatientset}[1]{\ensuremath{\mathbb{N}^{(#1)}_{d,p}}}

\newcommand*{\multgate}{\ensuremath{\gat_{\textnormal{Mult}}}}

\newcommand*{\crscgate}{\ensuremath{\gat_{\textnormal{CRS-C}}}}

\newcommand*{\ltgate}{\ensuremath{\gat_{\textnormal{LT}}}}

\newcommand*{\ufimultgate}{\ensuremath{\gat_{\textnormal{UFI-Mult}}}}

\newcommand*{\compgate}{\ensuremath{\gat_{\textnormal{Comp}}}}

\newcommand*{\aecselprot}{\ensuremath{\prot_{\textnormal{KEP-Rnd}}}}

\newcommand*{\aecselfunc}{\ensuremath{\func_{\textnormal{KEP-Rnd(\welfare)}}}}

\newcommand*{\multgatefunct}{\ensuremath{\gatfunc_{\textnormal{Mult}}}}

\newcommand*{\crscsinglegatefunct}{\ensuremath{\gatfunc_{\textnormal{CRS-C}}}}

\newcommand*{\compgatefunc}{\ensuremath{\gatfunc_{Comp}}}

\newcommand*{\ltgatefunct}{\ensuremath{\gatfunc_{\textnormal{LT}}}}

\AtBeginDocument{%
  \providecommand\BibTeX{{%
    \normalfont B\kern-0.5em{\scshape i\kern-0.25em b}\kern-0.8em\TeX}}}

\setcopyright{none}
\renewcommand\footnotetextcopyrightpermission[1]{}
\begin{document}
\fancyhead{}

\title{A Privacy-Preserving Protocol for the Kidney Exchange Problem}

\author{Malte Breuer}
\affiliation{%
  \institution{RWTH Aachen University}
%  \streetaddress{P.O. Box 1212}
   \city{Aachen}
   \state{Germany}
}
\email{breuer@itsec.rwth-aachen.de}

\author{Ulrike Meyer}
\affiliation{%
	\institution{RWTH Aachen University}
	%  \streetaddress{P.O. Box 1212}
	   \city{Aachen}
	   \state{Germany}
}
\email{meyer@itsec.rwth-aachen.de}

\author{Susanne Wetzel}
\affiliation{%
	\institution{Stevens Institute of Technology}
	%  \streetaddress{P.O. Box 1212}
	   \city{Hoboken}
	   \state{NJ, USA}
}
\email{swetzel@stevens.edu}

\author{Anja Mühlfeld}
\affiliation{%
	\institution{University Hospital RWTH Aachen}
	%  \streetaddress{P.O. Box 1212}
	   \city{Aachen}
	   \state{Germany}
}
\email{amuehlfeld@ukaachen.de}

%\renewcommand{\shortauthors}{}

\begin{abstract}
  Kidney donations from living donors form an attractive alternative to long waiting times on a list for a post-mortem donation. However, even if a living donor for a given patient is found, the donor's kidney might not meet the patient's medical requirements. If several patients are in this position, they may be able to exchange donors in a cyclic fashion. Current algorithmic approaches for determining such exchange cycles neglect the privacy requirements of donors and patients as they require their medical data to be centrally collected and evaluated. 
In this paper, we present the first distributed privacy-preserving protocol for kidney exchange that ensures the correct computing of the exchange cycles while at the same time protecting the privacy of the patients' sensitive medical data. We prove correctness and security of the new protocol and evaluate its practical performance.

\end{abstract}





\maketitle
\copyrightnotice{\copyright\space 2020 Association for Computing Machinery. This is the author's version of the work. It is posted here for your personal use. Not for redistribution. The definitive version was published in \emph{19th Workshop on Privacy in the Electronic Society (WPES'20), November 9, 2020, Virtual Event}, https://doi.org/10.1145/3411497.3420213}
\section{Introduction}\label{sec:introduction}

According to Eurotransplant's 2019 Annual Report~\cite{EurotransplantAnnualReport2019}, at the end of 2019 a total of 10,723 patients were on the waiting list for a kidney transplant from a post mortem donation in the participating countries.\footnote{Austria, Belgium, Croatia, Germany, Hungary, Luxemburg, the Netherlands, and Slovenia.} Patients in need of a kidney transplant can considerably reduce their waiting time if they find a compatible living donor. Such a living donor typically is a person with strong personal ties to the patient such that she is willing to donate one of her kidneys to the patient. While living donations increase the number of patients receiving a kidney transplant, only about 37\% of all kidneys transplanted in 2019 in the Eurotransplant region corresponded to such living donations~\cite{EurotransplantAnnualReport2019}. Although in many cases a patient can find a willing donor, this donor's kidney is often incompatible with the patient's medical characteristics. 

\begin{figure*}[t]
	\includegraphics[width=0.78\linewidth]{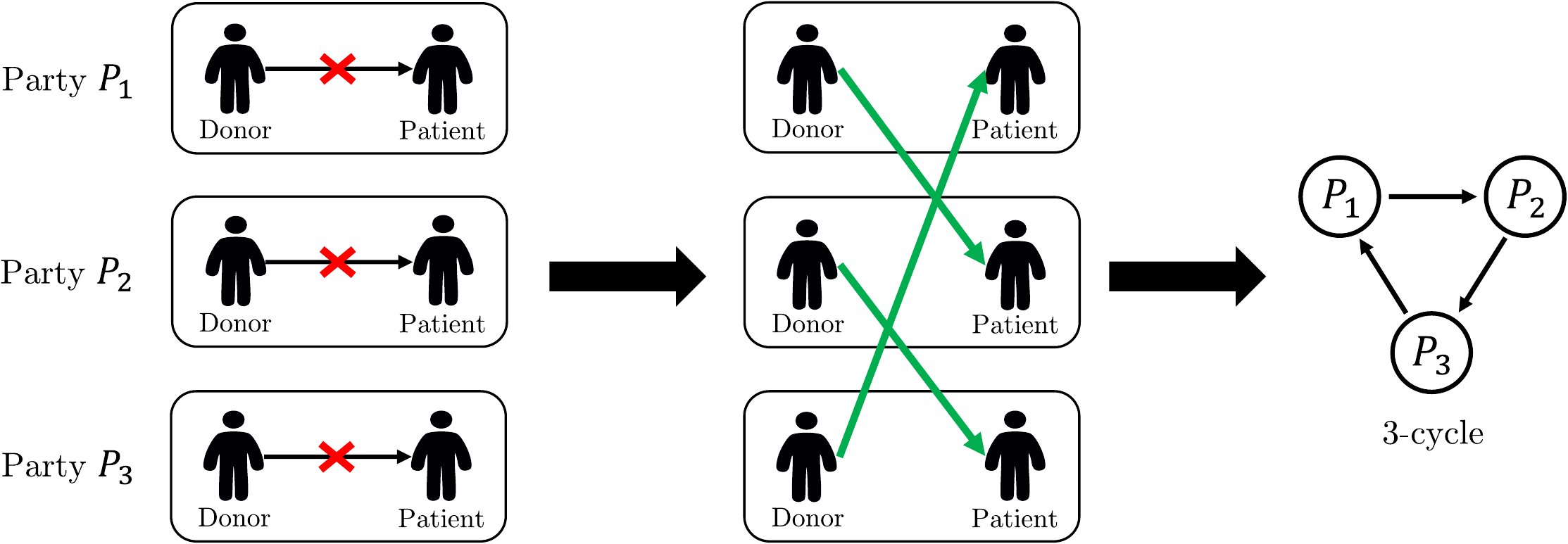}
	\caption{Living donor exchange based on an exchange cycle of size~3 between three parties each consisting of a patient and her incompatible donor.}
	\label{fig:threeCycle}
\end{figure*}

To increase the number of living donations, many countries world-wide allow so-called living donor exchanges. Here, patients with incompatible living donors are allowed to exchange donors in a cyclic fashion such that each patient whose donor gives her kidney to another patient also receives a donation from some other patient's donor. All transplants in such a cycle are typically required to be executed simultaneously~\cite{AbrahamClearingAlgorithms2007}. This prevents that a donor withdraws from donating as soon as her patient received a kidney. Due to the large number of medical staff and the vast resources that are needed to carry out a kidney transplant, the number of simultaneously executable transplants is limited. Commonly, the maximum cycle size $ \cyclesize $ considered for kidney exchange is $ \cyclesize = 3 $ such that a maximum of 6 operations on three donors and patients each are carried out simultaneously~\cite{AbrahamClearingAlgorithms2007}. Figure~\ref{fig:threeCycle} shows an example of three \emph{parties}, i.e., pairs consisting of a patient and her incompatible donor who exchange their donors in form of an exchange cycle of size~$ 3 $.

Given a fixed set of parties, the problem of finding a set of disjoint exchange cycles that allows for as many patients as possible to receive a compatible kidney is known as the \emph{Kidney Exchange Problem} (KEP)~\cite{AbrahamClearingAlgorithms2007}. The problem can be formulated as a graph problem on the so-called \emph{compatibility graph} in which each party is represented as a node in the graph and a directed edge is added from node $ i $ to node $ j $ iff the donor of party~$ \partysymbol{i} $ is compatible with the patient of party~$ \partysymbol{j} $ (cf.~Figure~\ref{fig:threeCycle}). Solving the KEP then corresponds to finding a set of exchange cycles in the compatibility graph that maximizes the number of patients that can receive a kidney transplant.

While in the past many algorithmic approaches for efficiently solving the KEP have been proposed (e.g.,~\cite{AbrahamClearingAlgorithms2007,AndersonTSP2015,RothEfficientKidneyExchange2007}), these approaches do not consider the existence of an adversary trying to compromise the parties' privacy or to manipulate the computation of the exchanges. However, given the sensitivity of a context such as kidney exchange, algorithmic solutions should be resilient against any form of manipulation by an external adversary. To the best of our knowledge, the existing algorithmic solutions for kidney exchange require the parties to reveal their sensitive medical data to a central platform where all this data is gathered and the exchange cycles to be executed are determined. With the central storage of all data, an adversary only needs to compromise one entity in order to obtain access to the parties' medical data. If such an attack remains undetected, the adversary may even be able to manipulate the actual computation of the exchange. An adversary may thereby, e.g., wrongfully force that a particular patient receives a compatible donor organ. 

In this paper, we address this shortcoming by presenting the first privacy-preserving protocol for solving the KEP. Specifically, we propose a de-centralized approach that allows the parties to keep their medical data private at all times. We prove correctness and security of our protocol in the presence of a \emph{semi-honest} adversary where an adversary controls a fixed set of \emph{corrupted} parties that strictly follow the protocol specification but try to learn as much as possible about the honest parties' input. To this end, we use \emph{Secure Multi-Party Computation} (SMPC) which is a cryptographic primitive that allows a fixed set of parties to compute a functionality~$ \func $ without the need of a trusted central entity such that a party only knows its private input and learns the output of the computation and what can be deduced from both. Using the formalism of SMPC allows us to formally prove correctness and security of our new protocol. Note that security in the semi-honest model is sufficient to prevent any meaningful manipulation of the computed exchange as all computations are executed on encrypted data. Thereby, it is impossible for an adversary to adapt a patient's input in order to increase her chances of finding a compatible donor.

Our privacy-preserving kidney exchange protocol builds on an existing SMPC protocol for privacy-preserving multi-party bartering~\cite{WuellerSemiHonestBartering2017} which allows a set of parties to determine a trade that is optimal w.r.t.\ a pre-defined welfare function (e.g., maximizing the number of parties that can trade). In our protocol for kidney exchange, a party consists of a patient and her incompatible donor which we also refer to as a \emph{patient-donor pair}. The private input of each party comprises the medical data of their patient and donor that is necessary to determine the compatibility between a patient and a donor. At the beginning of our protocol, all parties compute an encrypted adjacency matrix encoding the compatibility graph induced by their private inputs. To this end, we introduce a new SMPC protocol for privacy-preserving compatibility check for kidney exchange. Upon input of the medical data relevant for a kidney exchange between a patient and a donor, this protocol outputs an encryption of~$ 0 $ if a transplant between the patient and the donor~is~not possible (based on their medical data) and an encryption of~$ 1 $ if the donor may be compatible with the patient. Note that it is not possible to determine the compatibility between a patient and a donor with certainty on an algorithmic basis as the final decision has to be made by medical experts. Then, the constructed encrypted adjacency matrix is evaluated against a pre-computed set of all existing exchange constellations. This set is generic w.r.t.\ the input of the patient-donor pairs, i.e., it comprises all possible constellations in which the parties could exchange their donors. Thereby, those exchange constellations that can be executed based on the parties' private inputs (i.e., their relevant medical data) are determined in an oblivious fashion. In particular, the parties do not learn which of the constellations are executable. Finally, one of the executable exchange constellations is chosen uniformly at random such that the number of kidney transplants that can be executed is maximized. Each party's output then comprises suggested exchange partners for that party.\footnote{Note that in practice it is necessary that a medical professional reviews and verifies the determined exchange cycles before the transplants may proceed.} 

\begin{figure*}[!t]
	\input{./figures/paillier}
	\caption{Overview of the Threshold Paillier Cryptosystem from~\cite{FouquePaillierThreshold2001}.}
	\label{fig:paillier}
\end{figure*}

The main difference between our privacy-preserving kidney exchange protocol and the protocol for privacy-preserving bartering from~\cite{WuellerSemiHonestBartering2017} is the construction of the encrypted adjacency matrix that encodes the compatibility graph. While in bartering, computing the compatibility between two parties only requires to check whether their respective offers and demands are the same, in kidney exchange the compatibility check involves the comparison of complex medical data of the donor and the patient of the respective two parties. Thus, the main contribution of this paper is the development of a privacy-preserving compatibility check protocol for kidney exchange and its secure integration into the existing privacy-preserving bartering protocol. Furthermore, we formally prove the correctness and security of our new protocol for compatibility check and show that our modifications to the protocol for bartering for the context of kidney exchange do not impact the correctness and security of the original protocol.

We have implemented our protocol on top of an existing framework for SMPC~\cite{SMCMuseFramework2012} and report on the evaluation of its performance. Specifically, we have measured the runtime and induced network traffic for different numbers of parties and analyzed the performance impact of each of the different phases of our protocol.\\
\\
\noindent{\bf Outline:} The remainder of the paper is organized as follows: We first introduce notation, review existing building blocks (Section 2) and discuss related work (Section 3). In Section 4, we introduce our new privacy-preserving compatibility check for kidney exchange. Then, in Sections~5 and~6 we present and analyze our privacy-preserving kidney exchange protocol. We close this paper with some remarks on future work.

\section{Preliminaries and Notations}\label{sec:preliminaries}

Our new SMPC protocols make use of a threshold variant of the Paillier cryptosystem~\cite{FouquePaillierThreshold2001} and several previously introduced building blocks. In this section, we review these together with the notation we use in the remainder of this paper. 

For a finite set $ S $, $ r \rdraw S $ denotes that $ r $ is drawn uniformly at random from~$ S $. $ \natz $ is the set of natural numbers including $ 0 $ and $ \nats $ the set of natural numbers without $ 0 $. For $i\in \nats$, we define $ \natsbnd{i} := \{1,...,i\}$. The Iverson-Bracket $ \iverson{\cdot} $ for a logical statement~$ B $ is defined as $ \iverson{B} := 1 $ iff $ B $ is true and $ \iverson{B} := 0 $, otherwise. We denote the set of all prime numbers contained in an integer interval $ \mathcal{I} $ by $ \primeset_\mathcal{I} $. By $ \parties := \{1, ..., \numparties\} $ we denote the index set of the parties $ \partysymbol{1},..., \partysymbol{\numparties}$ participating in a multi-party protocol $\prot$. $ \corparties $ refers to the index set of the corrupted parties.

\subsection{Threshold Paillier Cryptosystem}\label{sub:pre_homEnc}

The SMPC protocols presented in this paper are based on the  $ (\partythresh, \numparties) $ threshold variant of the Paillier cryptosystem introduced in~\cite{FouquePaillierThreshold2001}, which is an additively homomorphic cryptosystem providing for semantic security against chosen-plaintext attacks. The decryption key is distributed among all $ \numparties $ parties such that at least $ \partythresh \leq \numparties $ parties have to collaborate to decrypt a ciphertext. Figure~\ref{fig:paillier} shows an overview of the key generation, the encryption function, and the homomorphic properties of the threshold Paillier variant we use in this paper.

For convenience, we omit the public and private key from notation and simply write $ \enc{m} := E(m) $ to denote the encryption of a message $ m $. We denote the entrywise encryption of a vector \linebreak $ U = (u_1{,}..., u_n) $ by $ \enc{U} := (\enc{u_1}{,}..., \enc{u_n}) $ and the entrywise encryption of a matrix $ A := (a_{1,1}{,}..., a_{m,n}) $ by $ \enc{A} := (\enc{a_{1,1}}{,}..., \enc{a_{m, n}}) $. Furthermore, we define $ \enc{U[i]} := \enc{u_i} $ and $ \enc{A[i,j]} := \enc{a_{i,j}} $.

\subsection{Secure Multi-Party Computation}\label{sub:pre_smpc}

In SMPC a fixed set of parties $ \partysymbol{1}, ..., \partysymbol{\numparties} $ jointly compute an $ \numparties $-input functionality $ \func : (\{0, 1\}^*)^\numparties \rightarrow (\{0, 1\}^*)^\numparties $ such that no party learns anything more than its private input, the computed output, and what can be deduced from both. This still holds in the presence of an adversary controlling $ \partythresh < \numparties $ parties. The goal of the adversary is to gather as much knowledge as possible on the honest parties' input or to manipulate the outcome of the computation.

In this paper, we consider a \emph{semi-honest} adversary, i.e., the parties controlled by the adversary follow the protocol specifications but try to learn as much as possible on the honest parties' input. The semi-honest adversary model is sufficient for many real-world applications~\cite{KolesnikovSecureTwoParty2006} where the protocol is embedded into complex systems. In such settings, the benefit from deviating from the protocol specification is typically small compared to the effort it takes~\cite{GoldreichFoundationsTwo2004}. Besides, deriving a protocol secure in the semi-honest model is often a first step towards a protocol secure in the presence of a malicious adversary where the parties may arbitrarily deviate from the protocol specification. 

Security in the semi-honest model is defined as follows. The \emph{view} of a party $ \partysymbol{i} $ during the execution of a multi-party protocol~$ \prot $ with input $ \overline{x} := (x_1, ..., x_\numparties) $ and security parameter $ \secparam $ is denoted by $\text{VIEW}^\prot_i(\overline{x}, \secparam) := (x_i, r_i, m_{i, 1}, ..., m_{i, k})$ with $ m_{i, j} $ being the $ j $-th message received by party~$ \partysymbol{i} $ during the execution of $ \prot $ and $ r_i $ representing the internal coin tosses of party~$ \partysymbol{i} $. Let $\text{OUTPUT}_i^\prot(\overline{x}, \secparam)$ refer to the output of party~$ \partysymbol{i} $.
We use $\overline{x}_{\corparties}$, $\func_{\corparties}(\overline{x})$, and $\text{VIEW}_{\corparties}^\prot(\overline{x}, \secparam)$ as short-hand notation for $(x_{i_1}, ..., x_{\numcorparties_i})$, $(\func_{i_1}(\overline{x}), ..., \func_{i_\numcorparties}(\overline{x}))$, and \linebreak $(\corparties, \text{VIEW}_{i_1}^\prot(\overline{x}, \secparam), ..., \text{VIEW}_{i_\numcorparties}^\prot(\overline{x}, \secparam))$.

\begin{definition}[Security in the Semi-Honest Model \cite{GoldreichFoundationsTwo2004}] \label{defn:semiHonest}
		A multi-party protocol $\prot$ securely computes a functionality $\mathcal{F}$ if there exists a probabilistic polynomial time algorithm $ \simu $ such that for every $ \numcorparties $ with $0 < \numcorparties < \numparties$ it holds that $ \{(\simu(1^\secparam, \corparties, \overline{x}_{\corparties}, \func_{\corparties}(\overline{x})), \func(\overline{x}))\}_{\overline{x}, \secparam} $ and $ \{(\textnormal{VIEW}_{\corparties}^{\prot}(\overline{x}, \secparam), \textnormal{OUTPUT}^{\prot}(\overline{x}, \secparam))\}_{\overline{x}, \secparam} $ are computationally indistinguishable.
\end{definition}

We call $ \simu $ the \emph{simulator} and denote the simulated values by angle brackets~$ \simval{\cdot} $. We distinguish between protocol functionalities~$ \func $ (resp., protocols~$ \prot $) and gate functionalities~$ \gatfunc $ (resp., gates~$ \gat $). A gate functionality~$ \gatfunc $ (resp., gate~$ \gat $) is a protocol that receives encrypted inputs and/or returns encrypted outputs.

\subsection{Existing Gates}\label{sub:pre_gates}

We review three existing gate functionalities which we use in our newly developed protocols. The complexities of their implementations are given in Table~\ref{tab:complexities}.

\begin{definition}[$ \multgatefunct $: Secure Multiplication \cite{Cramer_MultiParty_2001}]
		Let each party $ \partysymbol{i} \ (\forall i \in \parties) $ hold the two encrypted integers $ \enc{x} $ and $ \enc{y} $. Then, gate functionality $ \multgatefunct $ is specified as $ \enc{x \cdot y} \leftarrow \multgatefunct(\enc{x}, \enc{y}) $.
\end{definition}

As shorthand notation for the subsequent execution of gate $ \multgate $ we use $ \ufimultgate $ (Unbound Fan In Multiplication) as introduced in~\cite{WuellerDiss2018}.

\begin{definition}[$ \ltgatefunct $: Secure Less-Than Comparison \cite{WuellerDiss2018}]
		Let each party $ \partysymbol{i} \ (\forall i \in \parties) $ hold the two encrypted integers $ \enc{x} $ and $ \enc{y} $. Then, gate functionality $ \ltgatefunct $ is given as $ \enc{\iverson{x < y}} \leftarrow \ltgatefunct((\enc{x}, \enc{y})) $.
\end{definition}

In~\cite{WuellerDiss2018} the author presents a gate for secure comparison in the semi-honest model that provides for shared output, i.e., each party receives a bit such that the XOR of all these bits yields the output bit $ b $. However, it is trivial to modify this gate such that it provides each party with an encrypted output bit $ \enc{b} $ without revealing any information on $ b $ and thus implementing functionality~$ \ltgatefunct $.

\begin{definition}[$ \crscsinglegatefunct $: Conditional Random Selection with Output \linebreak Check \cite{Wueller_CRSmulti_2019}]
		Let each party $ \partysymbol{i} $ $ (\forall i \in \parties) $ hold two vectors $ \enc{U} $ and $ \enc{V} $ of length~$ n $. Let $ \enc{U} $ be an encrypted indicator vector and let $ \enc{V} $ be an encrypted value vector. Then functionality $ \crscsinglegatefunct $ is given as $ ((\enc{u_k^*}, \enc{v_k^*})) \leftarrow \crscsinglegatefunct((\enc{U}, \enc{V})) $ with $ \enc{u_k^*} := \blind{\enc{u_k}} $ and $ \enc{v_k^*} := \blind{v_k} $ where $ k \rdraw \{l \in \natsbnd{n}: \ u_l = \max(u_1, ..., u_n)\} $ if there is at least one $ l \in \natsbnd{n} $ such that $ u_l > 0 $. Otherwise, $ \crscsinglegatefunct $ returns $ (\emptystr, \emptystr) $ where $ \emptystr $ denotes the empty string.
\end{definition}

\begin{table}[!t]
	\centering
\begin{tabular}{lcc}
	Gate			& Communication complexity 				& Round complexity 			\\ \midrule
	$ \multgate $	& $ \bigo{\numparties \secparam} $		& $ \bigo{1} $				\\
	$ \ltgate $		& $ \bigo{\numparties \secparam} $		& $ \bigo{\numparties} $	\\
	$ \crscgate $	& $ \bigo{\numparties n \secparam} $	& $ \bigo{\numparties n} $	\\	
\end{tabular}
	\caption{Complexities of gate implementations~\cite{Cramer_MultiParty_2001,WuellerDiss2018,Wueller_CRSmulti_2019}.}
	\label{tab:complexities}
\end{table}
\section{Related Work}\label{sec:relatedWork}

Four main directions of work are related to the problem we solve: (1)~Conventional algorithms that solve the KEP without considering privacy. (2)~Privacy-preserving bartering protocols that can identify cyclic exchange options between parties bartering conventional goods while meeting specific privacy guarantees. (3)~Protocols for private intersection cardinality testing that solve a problem that is similar to a subproblem we face in our privacy-preserving protocol for compatibility check in kidney exchange~(cf.~Section~\ref{sub:comp_prot}). (4)~Privacy-preserving protocols for matching in bipartite graphs.

\subsection{Conventional Static Kidney Exchange}
Currently, the most efficient solutions for the KEP are based on Integer-Programming (IP) techniques (e.g., ~\cite{AbrahamClearingAlgorithms2007,AndersonTSP2015,RothEfficientKidneyExchange2007}). All these approaches consider a graph where each patient-donor pair corresponds to a node and an edge is added between two pairs iff the donor of the first pair is compatible with the patient of the second pair. The most efficient approach to date is referred to as the \emph{cycle formulation} where a binary decision variable is introduced for each cycle up to a cycle size bound $ \cyclesize $. The constraint is that each node is part of at most one cycle and the goal is to maximize the number of nodes in the solution. While these approaches solve the KEP efficiently in the non-privacy-preserving setting, they cannot be used as a basis for a privacy-preserving SMPC protocol since not all optimal solutions are obtained after the same number of optimization rounds. Instead the number of rounds needed depends on the input. Yet, a privacy-preserving protocol requires a control flow which is independent of the input. Thus, for our privacy-preserving kidney exchange protocol we have to take a different approach.

\subsection{Privacy-Preserving Multi-Party Bartering}
In privacy-preserving multi-party bartering, the parties strive to determine trades such that their offers and demands are matched while keeping their input private. Kidney exchange can be considered as a special case of bartering where a party consists of a patient-donor pair, the offer corresponds to the donor's medical data that is relevant for the kidney exchange, and the demand corresponds to the corresponding patient's medical data.
Frikken and Opyrchal~\cite{Frikken_TwoPartyPPBartering_2008} propose a two-party SMPC protocol to compute a trade from which both parties benefit allowing the parties to keep their utility on a public set of commodities private.
Kannan et al.~\cite{KannanParetoOptimalExchange15} present a protocol that computes trades such that the parties can keep their preferences over a public set of commodities as well as their offered commodity private.
In contrast to these two approaches, in the case of a privacy-preserving kidney exchange protocol, the set of commodities, which corresponds to the patients' and donors' medical data, has to remain private.
Wüller et al.~\cite{WuellerDiss2018,WuellerTradeChainDetection2018,WuellerMaliciousBartering2017,WuellerSemiHonestBartering2017,WuellerHungarianBartering2017} propose several SMPC protocols for multi-party bartering that vary w.r.t.\ the level of security they provide, the trade structures that can be computed, and their performance. All protocols exhibit the property that during their execution a party only knows its private input and learns the computed output and what can be deduced from both. In particular, the protocol from~\cite{WuellerSemiHonestBartering2017} allows for the computation of trades consisting of cycles of bounded size and it provides for security in the semi-honest model. As these are properties we also require for a privacy-preserving kidney exchange, we use this protocol as the basis for our privacy-preserving kidney exchange protocol (cf.~Section~\ref{sub:prot_specification}).

\subsection{Private Intersection Cardinality Testing}
Private intersection cardinality testing~(PICT) refers to the problem of testing whether the intersection between sets of different parties is larger than a threshold~$ T $. Existing protocols that solve the PICT problem (e.g.,~\cite{Badrinarayanan_ThresholdPSI_2020,Freedman_PrivateMatching_2004,Ghosh_ThresholdPSI_2019}) are either only designed for the two-party case or for the multi-party case where each party inputs a private set and it is checked whether the size of the intersection of all these sets is larger than~$ T $. However, the privacy-preserving compatibility check for kidney exchange~(cf.~Section~\ref{sub:comp_prot}) requires a multi-party protocol that computes whether the intersection between the sets of only two parties is empty (resp., non-empty). Furthermore, the existing protocols for PICT assume that the size of the input sets is public knowledge whereas in the case of kidney exchange the actual size of the sets has to remain private. We therefore devise a new approach for computing PICT that meets the special requirements of the compatibility check for kidney exchange~(cf.~Section~\ref{sub:comp_prot}).

\subsection{Privacy-Preserving Matching in Bipartite Graphs} 
In the literature there are several protocols for the privacy-preserving computation of a matching in a bipartite graph (e.g.,~\cite{Blanton_ObliviousBipartiteMatching_2015,Golle_PrivateBipartiteMatching_2006,WuellerHungarianBartering2017}). While these approaches could be used to solve the sub-problem of the KEP that restricts itself to finding only cycles of size 2 or to finding cycles of unbounded size (cf.~\cite{WuellerHungarianBartering2017}), these protocols do not support the determining of exchange cycles of bounded size larger than 2. Thus, they cannot be used in the context of the KEP when considering a maximum cycle size~$ \cyclesize = 3 $.

\section{Privacy-Preserving Compatibility Check for Kidney Exchange}\label{sec:comp}

Our new gate~$ \compgate $ computes the compatibility between a patient in need of a donor organ and a potential donor in a privacy-preserving fashion. Note that the gate~$ \compgate $ is the main new building block that is required to modify the existing SMPC protocol for privacy-preserving multi-party bartering from~\cite{WuellerSemiHonestBartering2017} such that it can be used to solve the KEP. Before providing the detailed gate specification (cf.\ Section~\ref{sub:comp_prot}), we discuss the relevant medical data each party has to provide as input (cf.\ Section~\ref{sub:comp_medData}).

\begin{table}[!t]
	\centering
\begin{tabular}{c c c}
	Blood Type \ \  	& Can Donate To \ \	& Can Receive From  	\\ \midrule
	O				& O, B, A, AB		& O						\\
	B				& B, AB				& O, B					\\
	A				& A, AB				& O, A					\\
	AB				& AB				& O, B, A, AB			\\
\end{tabular}
	\caption{Blood type compatibility for kidney exchange~\cite{EurotransplantManual2020}.}
	\label{tab:bloodComp}
\end{table}

\subsection{Relevant Medical Data}\label{sub:comp_medData}

To recall, the starting point for our gates and protocols is the fact that for for all parties~$ \partysymbol{i} $ ($ i \in \parties $), their respective donor and patient are not compatible and the goal is to find other parties such that compatibility between donors and patients from different parties potentially allows for a kidney transplant. Determining whether or not a patient and a donor are compatible requires the considering of various criteria. And even if all such criteria are met, this is not a guarantee that a transplant can be executed between what was determined to be a compatible patient and donor. In fact, the final decision as to whether a transplant should be carried out, lies with an experienced medical professional\footnote{Our kidney exchange protocol (cf.\ Section~\ref{sec:protocol}) outputs potential exchange partners for a patient-donor pair which will be recommended to and verified by medical professionals.}. Consequently, our compatibility check gate is designed to include only such checks of criteria that in case they do not match between a donor and a patient render a transplant impossible. According to transplant experts from the RWTH University hospital (which is a major transplant center in Europe), two such criteria are the blood type compatibility and HLA-type compatibility. However, it is important to note that our gate can be easily extended to include additional criteria if deemed necessary and suitable. 

\begin{table}[!t]
	\setlength{\tabcolsep}{4pt}
\renewcommand{\arraystretch}{1}
\small{
\centering
\begin{tabular}{ cc | cccc | c | c | cc}
	\multicolumn{2}{c}{\small HLA-A}	& \multicolumn{4}{c}{\small HLA-B}						& \multicolumn{1}{c}{\small HLA-C} & \multicolumn{1}{c}{\small HLA-DQ} 	& \multicolumn{2}{c}{\small HLA-DR} 	\\ \midrule
	A1			& A32	& B7		& B40		& B52		& B81		& C1		 				& DQ2							& DR1		& DR15				\\
	A2			& A33	& B8		& B41		& B53		& B82		& C2						& DQ3							& DR3		& DR16				\\
	A3			& A34	& B13		& B42		& B54		& 			& C3		 				& DQ4							& DR4		& 					\\
	A11			& A36	& B14		& B44		& B55		& 			& C4		 				& DQ5							& DR7		& 					\\
	A23			& A43	& B15		& B45		& B56		& 			& C5		 				& DQ6							& DR8		& 					\\
	A24			& A66	& B18		& B46		& B57		& 			& C6		 				& 								& DR9		& 					\\
	A25			& A68	& B27		& B47		& B58		& 			& C7		 				& 								& DR10		& 					\\
	A26			& A69	& B35		& B48		& B59		& 			& C8		 				& 								& DR11		& 					\\
	A29			& A74	& B37		& B49		& B67		& 			& 			 				& 								& DR12		& 					\\
	A30			& A80	& B38		& B50		& B73		&			&							&								& DR13		&					\\
	A31			&		& B39		& B51		& B78		&			&							&								& DR14		&					\\
\end{tabular}

}
	\caption{Antigens relevant for determining HLA-type compatibility in kidney exchange~\cite{KiefelHLAantigens2017}.}
	\label{tab:antigens}
\end{table}

Table~\ref{tab:bloodComp} summarizes what defines compatibility between the blood type of a donor and a patient. Specifically, there are four different blood types, i.e., O, A, B, AB and a donor with a certain blood type can only donate to patients with a certain blood type. For example, if the donor has blood type A, the patient's blood type has to be A or AB in order for the donor blood type compatibility to be met.

Table~\ref{tab:antigens} lists the antigens which are relevant in the context of a kidney transplant. Every human has certain antigens which are grouped into several HLA types. As there is an increased risk for acute rejection in case of an HLA incompatible kidney transplant~\cite{Kwon_HLAincompatible_2019}, our compatibility check only seeks for HLA compatible donors. Specifically, this means that the recipient has no HLA antibodies against the donor's antigens in the HLA-A, -B, -C, -DQ, and -DR loci.

Thus, our gate~$ \compgate $ computes an encrypted output bit~$ \enc{o} $ indicating the compatibility between a donor and a patient of two different parties based on their blood types and their antibodies/antigens. In particular, if $ o = 0 $, then the donor and the patient are not compatible and a transplant between them is not possible. If $ o = 1 $, the donor and the patient may be compatible and the final decision whether or not a transplant can be carried out lies with medical professionals~(cf.\ Footnote~3).

\subsection{Ideal Functionality and Gate Specification}\label{sub:comp_prot}

Before formally defining the functionality computed by our compatibility check gate for kidney exchange, we introduce the encoding of each party's input comprising the medical data described in Section~\ref{sub:comp_medData}. We encode compatibility with the patient's blood type by a binary indicator vector~$ \patientbloodvec{i} $ stating for each of the four existing blood types (O, B, A, AB) whether or not the patient of party~$ \partysymbol{i} $ can receive a kidney donor of that blood type. For example, if the patient of party~$ \partysymbol{i} $ has blood type B, the corresponding patient blood type indicator vector is $ \patientbloodvec{i} = [1, 1, 0, 0] $ (cf.\ Table~\ref{tab:bloodComp}). Analogously, compatibility with the donor's blood type is encoded by the donor blood type indicator vector $ \donorbloodvec{i} $ which indicates for each blood type whether the donor can donate to a patient with that blood type. The patient antibody vector~$ \antibodyvec{i} $ is also a binary indicator vector stating for each known antigen as specified in Table~\ref{tab:antigens} whether the patient has an antibody against it. Similarly, the donor antigen vector $ \antigenvec{i} $ indicates for each known antigen (cf.~Table~\ref{tab:antigens}) whether or not the donor has this antigen.

\begin{definition}[$ \compgatefunc $: Compatibility Check for Kidney Exchange]\label{defn:compCheck}
	Let a party $ \partysymbol{i} $ with $ i \in \parties $ hold blood type vector $ \donorbloodvec{i} $ and antigen vector $ \antigenvec{i} $ for its donor and another party $ \partysymbol{j} $ with $ j \in \parties $ hold blood type vector $ \patientbloodvec{j} $ and antibody vector $ \antibodyvec{j} $ for its patient. Then, gate functionality $ \compgatefunc $ is given as $ \enc{\out} \leftarrow \compgatefunc((\donorbloodvec{i}, \antigenvec{i}), (\patientbloodvec{j}, \antibodyvec{j})) $ where $ \enc{\out} $ is an encrypted bit indicating whether a donation of party $ \partysymbol{i} $'s donor to party $ \partysymbol{j} $'s patient can be excluded, i.e., $ \out = 0 $ iff party~$ \partysymbol{i} $'s donor is incompatible with party~$ \partysymbol{j} $'s patient.
\end{definition}

\begin{gate}[t]
	\caption{Secure Compatibility Check for Kidney Exchange}
	\label{prot:compCheck}
	\begin{myEnumerate}[leftmargin=0.35cm]
	\vspace*{0.6em}
	\item \underline{Input Sharing Phase}
	\vspace*{0.3em}
	\begin{myEnumerate}
		\item Party $ \partysymbol{i} $: Send $ \enc{\donorbloodvec{i}} $, $ \enc{\antigenvec{i}} $ to party $ \partysymbol{j} $
		\vspace*{0.3em}
		\item Party $ \partysymbol{j} $:
		\begin{myEnumerate}
			\item Set $ \enc{sum_B} := \enc{0} $, $ \enc{sum_A} := \enc{0} $
			\vspace*{0.3em}
			\item For $ k = 0 $ to $ \vert \patientbloodvec{j} \vert - 1 $:
			\begin{myEnumerate}
				\item If $ \iverson{\patientbloodvec{j}[k] = 1} $: Set $ \enc{sum_B} := \enc{sum_B} \addh \enc{\donorbloodvec{i}[k]} $
			\end{myEnumerate}
			\vspace*{0.3em}
			\item For $ k = 0 $ to $ \vert \antibodyvec{j} \vert - 1 $:
			\begin{myEnumerate}
				\item If $ \iverson{\antibodyvec{j}[k] = 1} $: Set $ \enc{sum_A} := \enc{sum_A} \addh \enc{\antigenvec{i}[k]} $
			\end{myEnumerate}
			\vspace*{0.3em}
			\item Broadcast $ \enc{sum_B} $, $ \enc{sum_A} $ to all parties
		\end{myEnumerate}
	\end{myEnumerate}
	\vspace*{0.6em}
	
	\item \underline{Compatibility Computation Phase}
	
	\vspace*{0.3em}
	\begin{myEnumerate}
		\item All parties:
		\begin{myEnumerate}
			\vspace*{0em}
			\item Jointly compute $ \enc{\out_B} \leftarrow \ltgate(\enc{0}, \enc{sum_B}) $
			\vspace*{0.3em}
			\item Jointly compute $ \enc{\out_A} \leftarrow \ltgate(\enc{sum_A}, \enc{1}) $
			\vspace*{0.3em}
			\item Jointly compute $ \enc{\out} \leftarrow \multgate(\enc{\out_B}, \enc{\out_A}) $
		\end{myEnumerate}
	\end{myEnumerate}
	\vspace*{0.6em}
	
	\item \underline{Output Phase}
	\vspace*{0.3em}
	\begin{myEnumerate}
		\item Party $ \partysymbol{i} $ outputs $ \enc{\out} $
	\end{myEnumerate}
\end{myEnumerate}

\end{gate}

Gate~\ref{prot:compCheck} implements gate functionality $ \compgatefunc $ (cf.~Definition~\ref{defn:compCheck}). In the following, we describe each of the three phases of gate $ \compgate $ in detail. 

\begin{enumerate}
	\item \emph{Input Sharing Phase}: 
	Party~$ \partysymbol{i} $ encrypts its donor blood type vector~$ \donorbloodvec{i} $ and antigen vector~$ \antigenvec{i} $ and sends both to party~$ \partysymbol{j} $. Based on the encrypted donor blood type indicator vector~$ \enc{\donorbloodvec{i}} $, party~$ \partysymbol{j} $ computes $ \enc{sum_B} $ which encodes the number of entries of the donor blood type indicator vector~$ \donorbloodvec{i} $ and the patient blood type indicator vector~$ \patientbloodvec{j} $ that are equal. To this end, it iterates over its patient blood type vector~$ \patientbloodvec{j} $ and increments $ \enc{sum_B} $ by the $ k $-th entry of the encrypted donor blood type vector~$ \enc{\donorbloodvec{i}} $ of party~$ \partysymbol{i} $ for all $ k $ which correspond to blood types that are compatible with the blood type of party $ \partysymbol{j} $'s patient, i.e., for all $ k $ where $ \patientbloodvec{j}[k] = 1 $. Thus, after Step~$ 1.2.2 $, $ \enc{sum_B} $ encodes the number of blood types to which the blood type of party~$ \partysymbol{i} $'s donor as well as the blood type of party~$ \partysymbol{j} $'s patient are compatible according to Table~\ref{tab:bloodComp}. Similarly, party~$ \partysymbol{j} $ iterates over its patient antibody vector~$ \antibodyvec{j} $ and computes $ \enc{sum_A} $ which encodes the number of antigens of party~$ \partysymbol{i} $'s donor against which party~$ \partysymbol{j} $'s patient has antibodies. Afterwards, party~$ \partysymbol{j} $ broadcasts the encrypted values $ \enc{sum_B} $ and $ \enc{sum_A} $ to all parties.
	
	\item \emph{Compatibility Computation Phase}:
	All parties jointly compute an encrypted bit~$ \enc{\out} $ indicating whether party~$ \partysymbol{i} $'s donor is incompatible with party~$ \partysymbol{j} $'s patient, i.e., if $ \out $ equals to~$ 0 $, party~$ \partysymbol{i} $'s donor cannot donate to party~$ \partysymbol{j} $'s patient. First, they check whether there is at least one blood type to which both the blood type of party~$ \partysymbol{i} $'s donor and the blood type of party~$ \partysymbol{j} $'s patient are compatible. To this end, the parties execute gate~$ \ltgate $ to obliviously determine whether $ sum_B $ is larger than $ 0 $ and store the result in the encrypted bit~$ \enc{\out_B} $. Afterwards, they verify that party~$ \partysymbol{j} $'s patient has no antibodies against party~$ \partysymbol{i} $'s donor. In particular, the parties execute gate~$ \ltgate $ to obliviously determine whether $ sum_A $ is less than $ 1 $ and store the result in the encrypted bit~$ \enc{\out_A} $. Finally, they execute a multiplication gate~$ \multgate $ to check whether both conditions hold and store the result in the encrypted bit~$ \enc{\out} $.
	
	\item \emph{Output Phase}: 
	Each party just outputs the encrypted bit~$ \enc{\out} $.
\end{enumerate}

In the following, we prove correctness and security of our compatibility check gate $ \compgate $ and analyze its complexity.

\paragraph{Correctness:}
To prove the correctness of gate~$ \compgate $, we show that after the computation of the input sharing phase $ sum_B $ is larger than $ 0 $ and $ sum_A $ is less than $ 1 $ iff the blood type of party~$ \partysymbol{i} $'s donor is compatible with the blood type of party~$ \partysymbol{j} $'s patient and at the same time party~$ \partysymbol{j} $'s patient has no antibodies against the antigens of party~$ \partysymbol{i} $'s donor. 
If the $ k $-th entry of the blood type vector~$ \patientbloodvec{j} $ equals~$ 1 $, i.e., party~$ \partysymbol{j} $'s patient is compatible with the blood type encoded by~$ \patientbloodvec{j}[k] $, $ sum_B $ is increased by the corresponding value of party~$ \partysymbol{i} $'s donor blood type indicator vector~$ \donorbloodvec{i} $. Thus, $ sum_B $ is only increased if both $ \patientbloodvec{j}[k] $ and $ \donorbloodvec{i}[k] $ equal $ 1 $ and, thus, only if the blood type of party~$ \partysymbol{i} $'s donor and the blood type of party~$ \partysymbol{j} $'s patient are compatible. Similarly, $ sum_A $ is only increased if both $ \antigenvec{i}[k] $ and $ \antibodyvec{j}[k] $ equal $ 1 $, i.e., only if party~$ \partysymbol{j} $'s patient has an antibody against the $ k $-th antigen of party~$ \partysymbol{i} $'s donor.
Thus, after the input sharing phase $ sum_B $ is larger than $ 0 $ iff the blood type of party~$ \partysymbol{i} $'s donor is compatible with the blood type of party~$ \partysymbol{j} $'s patient. Similarly, $ sum_A $ still equals $ 0 $, i.e., is less than $ 1 $, iff party~$ \partysymbol{j} $'s patient has no antibodies against any of the antigens of party~$ \partysymbol{i} $'s donor. 
Finally, assuming the correctness of gates~$ \ltgate $ and $ \multgate $, at the end of the compatibility computation phase $ \out $ equals~$ 1 $ iff $ sum_B $ is larger than $ 0 $ and $ sum_A $ is less than~$ 1 $. Otherwise, $ \out = 0$. Thus, gate~$ \compgate $ correctly computes gate functionality~$ \compgatefunc $ (cf.~Definition~\ref{defn:compCheck}).

\paragraph{Security:} We assume that all parties~$ \partysymbol{i} $ with $ i \in \corparties \subset \parties  $ and $ \corparties = \{i_{c_1}, ..., i_{c_\numcorparties}\} $ are corrupted. We have to differentiate between two cases for Step~$ 1.1 $. First, if $ i \in \corparties $, the simulator~$ \simu $ just encrypts party~$ \partysymbol{i} $'s input and sends it to party~$ \partysymbol{j} $. Second, if $ i \notin \corparties $ and $ j \in \corparties $, the simulator simulates each entry of the encrypted vectors $ \enc{\donorbloodvec{i}} $ and $ \enc{\antigenvec{i}} $ by random ciphertexts.
In the compatibility computation phase, the simulator can simulate the values $ \enc{sum_B} $ and $ \enc{sum_A} $ by $ \simval{\enc{sum_B}}~\rdraw~\cipherspace $ and $ \simval{\enc{sum_A}} \rdraw \cipherspace $, respectively. The calls to the gate $ \ltgate $ in Step~$ 2.1.1 $ and Step~$ 2.1.2 $ are simulated by calling the simulator of $ \ltgate $ on the input ($ \enc{0} $, $ \simval{\enc{sum_B}} $) and \linebreak ($ \simval{\enc{sum_A}} $, $ \enc{1} $), respectively. Similarly, the call to gate $ \multgate $ in Step~$ 2.1.3 $ is simulated by calling the simulator of $ \multgate $ on input ($ \simval{\enc{\out_B}} $, $ \simval{\enc{\out_A}} $). Finally, the simulator just outputs $ \simval{\enc{\out}} $. 

Due to the fact that the underlying cryptosystem is semantically secure, it follows that the simulated view is statistically indistinguishable from the real view.

\paragraph{Complexity:}
The input sharing phase only comprises operations that the parties $ \partysymbol{i} $ and $ \partysymbol{j} $ can execute locally. Thus, communication and round complexity of this phase are in $ \bigo{1} $. The compatibility computation phase requires the parties to execute gates $ \ltgate $ and $ \multgate $. Communication and round complexity of our implementation of $ \ltgate $ are $ \bigo{\numparties \secparam} $ and $ \bigo{\numparties} $, respectively, whereas the communication complexity of $ \multgate $ is $ \bigo{\numparties \secparam} $ and the round complexity is $ \bigo{1} $. Thus, gate $ \compgate $ overall exhibits a communication complexity of $ \bigo{\numparties \secparam} $ and a round complexity of $ \bigo{\numparties} $.

\section{Kidney Exchange Protocol}\label{sec:protocol}

In this section, we describe our privacy-preserving protocol~$ \aecselprot $ that solves the KEP for a fixed set of parties by choosing one optimal solution uniformly at random from the set of all optimal solutions. First, we introduce the necessary terminology together with the ideal functionality (Definition~\ref{defn:aecsel}) that is implemented by the protocol~$ \aecselprot $ (Section~\ref{sub:prot_overview}). Then, we provide a detailed specification of the protocol~$ \aecselprot $ together with an analysis of its correctness, security, and complexity~(Section~\ref{sub:prot_specification}).

\subsection{Terminology and Ideal Functionality}\label{sub:prot_overview}
The terminology introduced in this section is based on the terminology for privacy-preserving bartering from~\cite{WuellerMaliciousBartering2017} and adapted to the use case of kidney exchange.

Each party~$ \partysymbol{i} $ ($ \forall i \in \parties $) includes a patient seeking a kidney and her incompatible donor offering a kidney. The input of such a party then is the quote~$ \qu{i} = (\donor{i}, \patient{i}) $ where $ \donor{i} = (\donorbloodvec{i}, \antigenvec{i}) $ contains the donor-specific medical data that is relevant for a kidney transplant, i.e., the donor blood type vector~$ \donorbloodvec{i} $ and the donor antigen vector~$ \antigenvec{i} $, and $ \patient{i} = (\patientbloodvec{i}, \antibodyvec{i}) $ contains the patient-specific medical data that is relevant in the context of a kidney transplant, i.e., the patient blood type vector~$ \patientbloodvec{i} $ and the patient antibody vector~$ \antibodyvec{i} $. In our protocol, we only consider exchanges that are executed in form of so-called exchange cycles since this is the only way to ensure that a party's donor only donates her offered kidney iff the party's patient also receives a compatible kidney from the donor of another party. 

\begin{definition}[Exchange Cycle]\label{defn:cycle}
	For a set of parties $ \partysymbol{1}, ..., \partysymbol{\numparties} $ and the corresponding set of quotes $ \qu{1}, .., \qu{\numparties} $, an exchange cycle of size $ \cyclesize $ is a tuple $ (\partysymbol{i_1}, \partysymbol{i_2}, ..., \partysymbol{i_\cyclesize}) $ with $ i_l \neq i_k $ for $ l \neq k $ such that the donor of Party $ \partysymbol{i_l} $ is compatible with the patient of Party $ \partysymbol{i_{l+1}} $ for $ l \in \{1, ..., m-1\} $ and the donor of Party $ \partysymbol{i_m} $ is compatible with the patient of Party $ \partysymbol{i_1} $ according to Definition~\ref{defn:compCheck}.
\end{definition}

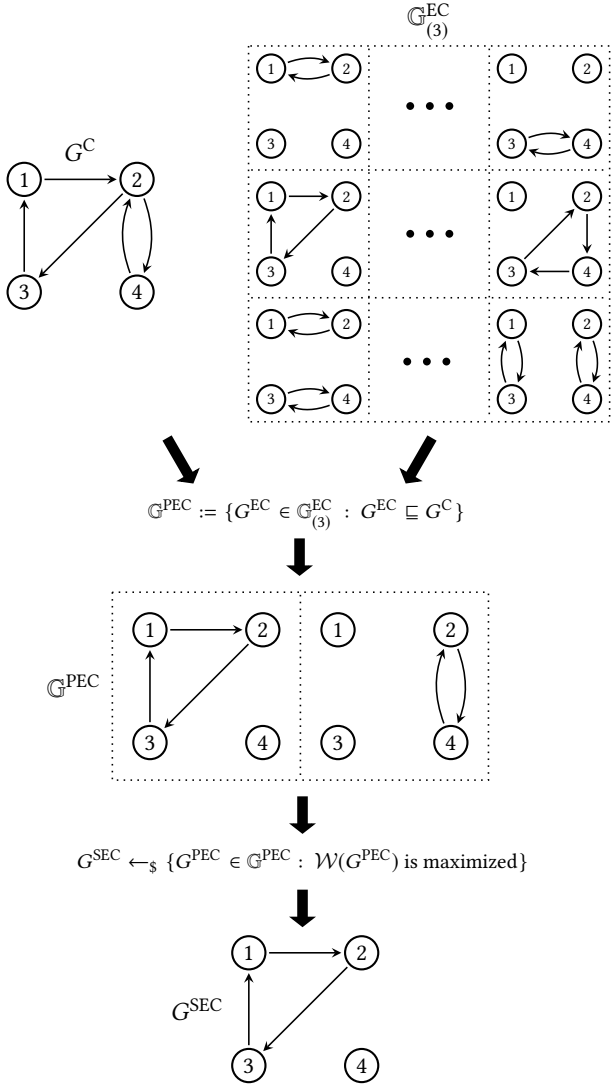
\begin{figure}[!t]
	\hspace*{0cm}
\vspace*{0cm}
\raisebox{1.5cm}{
\begin{tikzpicture}[ 
	> = stealth, % arrow head style
	shorten > = 1pt, % don't touch arrow head to node
	shorten < = 1pt,
	auto,
	node distance = 2cm, % distance between nodes
	semithick % line style
	]
	\tikzstyle{every state}=[
	draw = black,
	thick,
	fill = white,
	inner sep=2pt,
	minimum size = 1mm
	]
	
	\node[state] (v1) at (0,0) 				{$1$};
	\node[state] (v2) at (1.5,0) 			{$2$};
	\node[state] (v3) at (0,-1.5)			{$3$};
	\node[state] (v4) at (1.5,-1.5)		{$4$};
	
	\path[->] (v1) edge (v2);
	\path[->] (v2) edge (v3);
	\path[->] (v3) edge (v1);
	\path[->] (v2) edge[bend left = 20] (v4);
	\path[->] (v4) edge[bend left = 20] (v2);

	%title
	\node[align=center, yshift=0.5em] (title) at (current bounding box.north) {\large$ \cg $};

\end{tikzpicture}	
} 
\hspace*{1cm}
%\vspace*{-1.75cm}
\begin{tikzpicture}[ 
> = stealth, % arrow head style
shorten > = 1pt, % don't touch arrow head to node
shorten < = 1pt,
auto,
node distance = 2cm, % distance between nodes
semithick % line style
]
\tikzstyle{every state}=[
draw = black,
thick,
fill = white,
inner sep=2pt,
minimum size = 0.5mm
]

\draw[dotted] (0,0) rectangle (4.8,-5);
\draw[dotted] (0,-1.65) -- (4.8,-1.65);
\draw[dotted] (0,-3.35) -- (4.8,-3.35);
\draw[dotted] (1.6,0) -- (1.6,-5);
\draw[dotted] (3.2,0) -- (3.2,-5);

\node[align=center] (upperDots) at (2.37,-0.8) {\fontsize{28}{32} \selectfont ...};
\node[align=center] (middleDots) at (2.37,-2.5) {\fontsize{28}{32} \selectfont ...};
\node[align=center] (LowerDots) at (2.37,-4.2) {\fontsize{28}{32}\selectfont ...};

% UPPER LEFT
\node[state] (v1) at (0.3,-0.3) 		{\scriptsize$1$};
\node[state] (v2) at (1.3,-0.3) 		{\scriptsize$2$};
\node[state] (v3) at (0.3,-1.3)			{\scriptsize$3$};
\node[state] (v4) at (1.3,-1.3)			{\scriptsize$4$};

\path[->] (v1) edge[bend left = 20] (v2);
\path[->] (v2) edge[bend left = 20] (v1);

% UPPER RIGHT
\node[state] (v11) at (3.5,-0.3) 			{\scriptsize$1$};
\node[state] (v12) at (4.5,-0.3)			{\scriptsize$2$};
\node[state] (v13) at (3.5,-1.3)			{\scriptsize$3$};
\node[state] (v14) at (4.5,-1.3)		 	{\scriptsize$4$};

\path[->] (v13) edge[bend left = 20] (v14);
\path[->] (v14) edge[bend left = 20] (v13);

% MIDDLE LEFT
\node[state] (v6) at (0.3,-2) 			{\scriptsize$1$};
\node[state] (v7) at (1.3,-2) 			{\scriptsize$2$};
\node[state] (v8) at (0.3,-3)			{\scriptsize$3$};
\node[state] (v9) at (1.3,-3)		 	{\scriptsize$4$};

\path[->] (v6) edge (v7);
\path[->] (v7) edge (v8);
\path[->] (v8) edge (v6);

% MIDDLE RIGHT
\node[state] (v16) at (3.5,-2) 			{\scriptsize$1$};
\node[state] (v17) at (4.5,-2) 			{\scriptsize$2$};
\node[state] (v18) at (3.5,-3)			{\scriptsize$3$};
\node[state] (v19) at (4.5,-3)	 		{\scriptsize$4$};

\path[->] (v17) edge (v19);
\path[->] (v19) edge (v18);
\path[->] (v18) edge (v17);

% LOWER LEFT
\node[state] (v20) at (0.3,-3.7) 			{\scriptsize$1$};
\node[state] (v21) at (1.3,-3.7) 			{\scriptsize$2$};
\node[state] (v22) at (0.3,-4.7)			{\scriptsize$3$};
\node[state] (v23) at (1.3,-4.7)		 	{\scriptsize$4$};

\path[->] (v20) edge[bend left = 20] (v21);
\path[->] (v21) edge[bend left = 20] (v20);
\path[->] (v22) edge[bend left = 20] (v23);
\path[->] (v23) edge[bend left = 20] (v22);

% LOWER RIGHT
\node[state] (v24) at (3.5,-3.7) 			{\scriptsize$1$};
\node[state] (v25) at (4.5,-3.7) 			{\scriptsize$2$};
\node[state] (v26) at (3.5,-4.7)			{\scriptsize$3$};
\node[state] (v27) at (4.5,-4.7)	 		{\scriptsize$4$};

\path[->] (v24) edge[bend left = 20] (v26);
\path[->] (v26) edge[bend left = 20] (v24);
\path[->] (v25) edge[bend left = 20] (v27);
\path[->] (v27) edge[bend left = 20] (v25);

% TITLE
\node[align=center, yshift=1em, xshift=0em] (title) at (current bounding box.north) {\large $\ecsg_{(3)}$};
\end{tikzpicture}	

\vspace*{0.1cm}
\begin{tikzpicture}
\draw[-{Triangle[width=10pt,length=4pt]}, line width=4pt, rotate=30](0,0) -- (0, -0.7);
\end{tikzpicture}
\hspace*{2.5cm}
\begin{tikzpicture}
\draw[-{Triangle[width=10pt,length=4pt]}, line width=4pt, rotate=-30](0,0) -- (0, -0.7);
\end{tikzpicture}
\vspace*{0cm}

\begin{tikzpicture}
%\node[](left) at (0,0) {\small$\phantom{x} \pecg_i \sqsubseteq \cg $ $ \wedge $ $ \pecg_i \in \ecsg_{(3)} $ };
\node[](left) at (0,0) {\small$\phantom{x} \pecsg := \{\ecg \in \ecsg_{(3)} \ : \ \ecg \sqsubseteq \cg \} $ };
\end{tikzpicture}
\vspace*{-0.1cm}

\begin{tikzpicture}
\draw[-{Triangle[width=10pt,length=4pt]}, line width=4pt, rotate=0](0,0) -- (0, -0.5);
\end{tikzpicture}
\vspace*{0.1cm}

\hspace*{-1.1cm}
\begin{tikzpicture}[ 
		> = stealth, % arrow head style
		shorten > = 1pt, % don't touch arrow head to node
		shorten < = 1pt,
		auto,
		node distance = 2cm, % distance between nodes
		semithick % line style
	]
	\tikzstyle{every state}=[
	draw = black,
	thick,
	fill = white,
	inner sep=2pt,
	minimum size = 1mm
	]
	
%	\draw[dotted] (0,0.4) rectangle (5,-2.5);
%	\draw[dotted] (2.5,0.4) -- (2.5,-2.5);
	\draw[dotted] (0,0) rectangle (5,-2.5);
	\draw[dotted] (2.5,0) -- (2.5,-2.5);
	
	\node[state] (v1) at (0.5,-0.5) 			{$1$};
	\node[state] (v2) at (2,-0.5) 				{$2$};
	\node[state] (v3) at (0.5,-2)				{$3$};
	\node[state] (v4) at (2,-2)		 			{$4$};
	
	\path[->] (v1) edge (v2);
	\path[->] (v2) edge (v3);
	\path[->] (v3) edge (v1);
	
	\node[state] (v6) at (3,-0.5) 				{$1$};
	\node[state] (v7) at (4.5,-0.5)				{$2$};
	\node[state] (v8) at (3,-2)					{$3$};
	\node[state] (v9) at (4.5,-2)			 	{$4$};
	
	\path[->] (v7) edge[bend left = 20] (v9);
	\path[->] (v9) edge[bend left = 20] (v7);
	
	%\node[align=center] (pecg1) at (1.25,0) {$ \pecg_1 $};
	
	%\node[align=center] (pecg1) at (3.75,0) {$ \pecg_2 $};
	
	%title
	\node[align=center, yshift=0em, xshift=-0.5cm] (title) at (current bounding box.west) {\large$\pecsg$};
\end{tikzpicture}

\vspace*{0.1cm}
\begin{tikzpicture}
\draw[-{Triangle[width=10pt,length=4pt]}, line width=4pt](0,0) -- (0, -0.5);
\end{tikzpicture}

\begin{tikzpicture}
\node[](left) at (0,-0) {\small$ \aecg \rdraw \{\pecg \in \pecsg: \ \welfare(\pecg) \ \text{is maximized}\} $};
\end{tikzpicture}

\begin{tikzpicture}
\draw[-{Triangle[width=10pt,length=4pt]}, line width=4pt](0,0) -- (0, -0.5);
\end{tikzpicture}

\hspace*{-0.9cm}
\begin{tikzpicture}[ 
		> = stealth, % arrow head style
		shorten > = 1pt, % don't touch arrow head to node
		shorten < = 1pt,
		auto,
		node distance = 2cm, % distance between nodes
		semithick % line style
	]
	\tikzstyle{every state}=[
	draw = black,
	thick,
	fill = white,
	inner sep=2pt,
	minimum size = 1mm
	]
	
	\node[state] (v1) at (0,0) 			{$1$};
	\node[state] (v2) at (1.5,0) 		{$2$};	
	\node[state] (v3) at (0,-1.5)		{$3$};
	\node[state] (v4) at (1.5,-1.5)	 	{$4$};
			
	\path[->] (v1) edge (v2);
	\path[->] (v2) edge (v3);
	\path[->] (v3) edge (v1);
	
	%title
%	\node[align=center] (title) at (0.75, 0.5) {\large$\aecg$};
	\node[align=center, yshift=0em, xshift=-1.4em] (title) at (current bounding box.west) {\large$\aecg$};
	
\end{tikzpicture}	 
	\caption{Example of the privacy-preserving computation of the suggested exchange constellation graph~$ \aecg $ for a set of four parties according to functionality~$ \aecselfunc $, based on~\cite{WuellerMaliciousBartering2017}.}
	\label{fig:ke_notation}
\end{figure}

We call a set of exchange cycles \emph{disjoint} and \emph{simultaneously executable} if they have no party in common.  
The general idea of our approach for our privacy-preserving kidney exchange protocol is to first construct a \emph{compatibility graph} which reflects the exchanges which are possible for a given set of parties and their specific set of quotes. Then, we identify simultaneously executable exchange cycles within the compatibility graph to determine a set of exchange cycles that maximizes the number of parties that can receive a kidney transplant. Figure~\ref{fig:ke_notation} illustrates the relationships between the different graphs and how they are used to compute the functionality~$ \aecselfunc $ (cf.~Definition~\ref{defn:aecsel}).

\begin{definition}[$ \cg $: Compatibility Graph]\label{defn:cg}
	Given the private input quotes $ \qu{i} = (\donor{i}, \patient{i}) $ of all parties $ \partysymbol{i} $ with $ i \in \parties $, a compatibility graph $ \cg $ is a directed graph $ (V, E) $ with $ V := \parties = \{1, ..., \numparties\} $ and for any $ i, j \in \parties $ with $ i \neq j $ it holds that $ (i, j) \in E $ if the donor quote $ \donor{i} $ is compatible with the patient quote $ \patient{j} $ according to Definition~\ref{defn:compCheck}.
\end{definition}

The upper left corner of Figure~\ref{fig:ke_notation} depicts an example of a compatibility graph for four parties indicating the compatibility between those parties. Specifically, the edges of the compatibility graph encode the compatibility between the four parties based on the medical data of their donors and patients. For example, the edge $ (1,2) $ indicates that the donor of party~$ \partysymbol{1} $ is compatible with the patient of party~$ \partysymbol{2} $ according to Definition~\ref{defn:compCheck}.

In contrast to a compatibility graph, an \emph{exchange constellation graph} is generic, i.e., independent of a specific set of input quotes. Specifically, an exchange constellation graph represents one possible constellation of how the parties could exchange the kidneys of their respective incompatible donors.

\begin{definition}[$ \ecg $: Exchange Constellation Graph]\label{defn:ecg}
	An exchange constellation graph $ \ecg $ is a directed graph $ (V, E) $ with $ V := \parties $ where $ \forall i \in V: (\indeg{i} = \outdeg{i} = 1) \vee (\degree{i} = 0) $.
\end{definition}

While an exchange constellation graph only contains disjoint exchange cycles, a compatibility graph may also contain exchange cycles which are not simultaneously executable. The set of all exchange constellation graphs for a set of parties is denoted by~$ \ecsg $. If the set only consists of all exchange constellation graphs that contain exchange cycles up to size~$ m < \numparties $, we denote this by~$ \ecsg_{(m)} $. The upper right corner of Figure~\ref{fig:ke_notation} depicts the set~$ \ecsg_{(3)} $ on the example of four parties. This set includes all exchange constellation graphs containing a single cycle of size~$ \cyclesize = 2 $, all exchange constellation graphs containing a single cycle of size~$ \cyclesize = 3 $, and all exchange constellation graphs containing two cycles of size~$ \cyclesize = 2 $. 

\begin{definition}[Neighborhood Constellation]\label{defn:neighborhood}
	Given an exchange constellation graph $ \ecg = (V, E) $, for each node $ i \in V $ its neighborhood constellation is defined as $ \ndonorpatient{i} = (\ndonor{i}(\ecg), \npatient{i}(\ecg)) $. The source node of the incoming edge is referred to as the donating neighbor $ \ndonor{i}(\ecg) := j $ if $ \exists(j, i) \in E  $ and $ \ndonor{i}(\ecg) := 0 $, otherwise. Similarly, the target node of the outgoing edge is referred to as the receiving neighbor $ \npatient{i}(\ecg) := j $ if $ \exists(i, j) \in E $ and $ \npatient{i}(\ecg) := 0 $, otherwise.
\end{definition}

The set of all neighborhood constellations of a node $ i $ and a set of exchange constellation graphs $ \ecsg $ is denoted by $ \ndonorpatientset{i}(\ecsg) $.

The subgraph of a compatibility graph~$ \cg $ containing a set of simultaneously executable exchange cycles is referred to as a \emph{Potential Exchange Constellation} (PEC) \emph{graph}.

\begin{definition}[$ \pecg $: Potential Exchange Constellation Graph]\label{defn:pecg}
	Given $ \cg $, an exchange constellation graph $ \ecg $ is referred to as potential exchange constellation graph $ \pecg $ if $ \ecg $ is a subgraph of $ \cg $ (denoted by $ \ecg \sqsubseteq \cg $).
\end{definition}

$ \pecsg $ is the set of all potential exchange constellation graphs for a compatibility graph~$ \cg $ and a set of exchange constellation graphs~$ \ecsg $.

An example of the set~$ \pecsg $ for four parties is depicted in the middle part of Figure~\ref{fig:ke_notation}. The set contains the only two exchange constellation graphs from the set~$ \ecsg_{(3)} $ (upper right corner of Figure~\ref{fig:ke_notation}) that form a subgraph of the compatibility graph~$ \cg $ (upper left corner of Figure~\ref{fig:ke_notation}). 

The goal of our kidney exchange protocol is to determine a potential exchange constellation graph that is optimal w.r.t.\ a pre-defined welfare function.

\begin{definition}[Welfare Function]\label{defn:welfare}
	A welfare function $ \welfare(\cdot): \ecsg \rightarrow \natz $ maps an exchange constellation graph in $ \ecsg $ to a welfare $ w \in \natz $ that measures the overall utility of an exchange constellation graph.
\end{definition}

Throughout this paper we use the welfare function that counts the edges in an exchange constellation graph, i.e., that counts the number of patients that receive a kidney transplant in a given exchange constellation graph.

Based on the above introduced terminology, we define the ideal functionality~$ \aecselfunc $ which is implemented by our privacy-preserving protocol~$ \aecselprot $ for the KEP.

\begin{definition}[$ \aecselfunc $: Privacy-Preserving KEP with Random Selection]\label{defn:aecsel}
	Let each party $ \partysymbol{i} $ hold its private input quote $ \kequote{i} = (\donor{i}, \patient{i}) $ ($ \forall i \in \parties $) with $ \donor{i} = (\donorbloodvec{i}, \antigenvec{i}) $ being party $ \partysymbol{i} $'s donor input quote and $ \patient{i} = (\patientbloodvec{i}, \antibodyvec{i}) $ being party $ \partysymbol{i} $'s patient input quote. Further, let $ \ecsg $ be a publicly known set of exchange constellation graphs for $ \numparties $ parties and let $ \welfare(\cdot): \ecsg \leftarrow \natz $ be some publicly known welfare function. Then, functionality $ \aecselfunc $ is defined as 
	\begin{equation*}
		\left. \begin{array}{ll}
		(\out_1, ..., \out_\numparties)		&	\text{\textit{if }} \pecsg \neq \emptyset \\ \relax
		(0) 							& 	\text{\textit{otherwise}}
		\end{array} \right\} \leftarrow \aecselfunc(\qu{1}, ..., \qu{\numparties}, \ecsg)
	\end{equation*}
	with $ \aecg \rdraw \{\pecg \in \pecsg: \ \welfare(\pecg) \ \text{is maximized}\} $ and \linebreak $ \out_i = \ndonorpatient{i}(\aecg) $. 
\end{definition}

The exchange constellation graph that is determined by functionality~$ \aecselfunc $ is referred to as \emph{Suggested Exchange Constellation} (SEC) \emph{graph}. 
In particular, the graph~$ \aecg $ maximizes the number of patients that can receive a kidney transplant. If there is more than one graph maximizing the number of transplants, one of these is chosen uniformly at random as the suggested exchange constellation graph~$ \aecg $. Thus, the functionality~$ \aecselfunc $ goes beyond simply solving the KEP as it not only outputs some optimal solution but provides for \emph{unbiasedness} in that it outputs one optimal solution chosen uniformly at random from the set of all optimal solutions. The graph is called \emph{suggested} exchange constellation graph since the final decision of whether the computed kidney transplants can be carried out is still made by medical professionals. 

The bottom of Figure~\ref{fig:ke_notation}, depicts the suggested exchange constellation graph~$ \aecg $. In this example, the graph~$ \aecg $ suggests that three patients may be able to receive a kidney transplant and as such yields the maximum welfare of all potential exchange constellation graphs in the set~$ \pecsg $ (middle part of Figure~\ref{fig:ke_notation}). 

After introducing all necessary terminology, we now review the complete example of the computation of functionality~$ \aecselfunc $ between four parties as depicted in Figure~\ref{fig:ke_notation}.
The upper left corner depicts the compatibility graph~$ \cg $ indicating the compatibility between the four parties computed based on the medical data of the parties' patients and donors. The upper right corner of the figure shows the exchange constellation graph set~$ \ecsg_{(3)} $ for four parties and a maximum cycle size $ \cyclesize = 3 $. Recall that this set is pre-computed and does not depend on the specific inputs of the four patient-donor pairs that seek to compute functionality~$ \aecselfunc $. Based on the set of exchange constellation graphs~$ \ecsg_{(3)} $ and the compatibility graph~$ \cg $, the set of potential exchange constellation graphs~$ \pecsg $ is computed. This set forms a subset of the set of all exchange constellation graphs~$ \ecsg_{(3)} $ including only those exchange constellation graphs that form a subgraph of the compatibility graph (i.e., $ \forall \ecg \in \ecsg_{(3)} : \ecg \sqsubseteq \cg $). The set of all potential exchange constellation graphs~$ \pecsg $ is depicted in the middle part of Figure~\ref{fig:ke_notation}. Finally, one of those potential exchange constellation graphs that maximize the welfare function~$ \welfare $ (i.e., the number of patients that may receive a kidney transplant) is chosen uniformly at random as the suggested exchange constellation graph~$ \aecg $. In the example shown in Figure~\ref{fig:ke_notation}, the suggested exchange constellation graph~$ \aecg $ is the graph that contains one exchange cycle of size $ 3 $ including parties $ \partysymbol{1} $, $ \partysymbol{2} $, and $ \partysymbol{3} $. This graph yields the maximum welfare of the two potential exchange constellation graphs (middle part of Figure~\ref{fig:ke_notation}) since it may allow three patients to receive a kidney transplant whereas the other potential exchange constellation graph may only allow two patients to receive a kidney transplant. The graph~$ \aecg $ then defines the output of functionality~$ \aecselfunc $ for our example.

\subsection{Specification of Protocol $ \aecselprot $}\label{sub:prot_specification}

Our novel protocol~$ \aecselprot $ (cf.~Protocol~\ref{prot:aecsel}) implements functionality~$ \aecselfunc $ in the semi-honest model.
The input of each party~$ \partysymbol{i} $ to the protocol is the medical data of the party's donor and patient that is relevant for a kidney transplant (cf.~Definition~\ref{defn:compCheck}). In particular, this corresponds to the donor input $ \donor{i} = (\donorbloodvec{i}, \antigenvec{i}) $ containing the donor blood type vector~$ \donorbloodvec{i} $ and the donor antigen vector~$ \antigenvec{i} $ and the patient input $ \patient{i} = (\patientbloodvec{i}, \antibodyvec{i}) $ containing the patient blood type vector~$ \patientbloodvec{i} $ and the patient antibody vector~$ \antibodyvec{i} $. Recall that in our protocol the welfare function~$ \welfare(\cdot) $ corresponds to counting the number of patients that receive a kidney transplant, i.e., counting the number of edges in the exchange constellation graph.

\begin{protocol}[!t]
	\caption{$ \aecselprot $, based on \cite{WuellerSemiHonestBartering2017}.}
	\label{prot:aecsel}
	\small{
	\begin{myEnumerate}[leftmargin=0.35cm]
		\vspace*{0.6em}
		\item \underline{Construction Phase}	
		\vspace*{0.3em}
		\begin{myEnumerate}
			\item For each $ \partysymbol{i} \ (i \in \parties) $:
			
			\begin{myEnumerate}
				\item For each $ \partysymbol{j} \ (j \in \parties\setminus\{i\}): $
				
				\begin{myEnumerate}
					\item All parties jointly compute $ \enc{a_{i, j}} \leftarrow \compgate(\partysymbol{i}, \partysymbol{j}) $
				\end{myEnumerate}
			\end{myEnumerate}
		\end{myEnumerate}
		
		\vspace*{0.4em}
		\item \underline{Evaluation Phase}
		\vspace*{0.3em}
		\begin{myEnumerate}
			\item For each $ \ecg_k = (V, E) \in \ecsg_{(3)} $ with $ (i_l, j_l) \in E $, $ k \in \nats_{\vert \ecsg_{(3)} \vert} $, $ l \in \nats_{\vert E \vert} $:
			
			\begin{myEnumerate}
				\item All parties jointly compute \\
				\centering $ \enc{e_k} \leftarrow \ufimultgate((\enc{a_{i_1, j_1}}, ..., \enc{a_{i_{\vert E \vert}, j_{\vert E \vert}}})) $
				
			\end{myEnumerate}
			\vspace*{0.4em}
			\item All parties set $ \enc{L} := (\enc{e_1}, ..., \enc{e_{\vert \ecsg \vert}}) $
		\end{myEnumerate}
		
		\vspace*{0.4em}	
		\item \underline{Prioritization Phase}
		\vspace*{0.3em}
		\begin{myEnumerate}
			\item Each party locally computes \\ 
			\centering $ \enc{L_1} := (\enc{e_1} \crossh \welfare(\ecg_1), ..., \enc{e_{\vert \ecsg_{(3)} \vert}} \crossh \welfare(\ecg_{\vert \ecsg_{(3)} \vert})) $
		\end{myEnumerate}
		
		\vspace*{0.4em}	
		\item \underline{Mapping Phase}
		\vspace*{0.3em}
		\begin{myEnumerate}
			\item Each party $ \partysymbol{i} \ (i \in \parties) $:
			
			\begin{myEnumerate}
				\item Set $ S^{(i)} := \emptyset $
				\vspace*{0.4em}
				\item For each $ \ndonorpatient{i} \in \ndonorpatientset{i}(\ecsg_{(3)}): $
				
				\begin{myEnumerate}
					\vspace*{0.4em}
					\item Select $ p \rdraw \primeset_{I^{(i)}} \setminus S^{(i)} $
					\vspace*{0.4em}
					\item Update $ S^{(i)} = S^{(i)} \cup \{p^{(i)}_{\ndonorpatient{i}}\} $
				\end{myEnumerate}
			\end{myEnumerate}
			
			\item Party $ \partysymbol{\numparties}: $
			
			\begin{myEnumerate}
				\item Set 
				$ \enc{u_k^{(\numparties)}} := \enc{p^{(\numparties)}_{\ndonorpatient{\numparties}(\ecg_k)}} $ 
				with $ \ecg_k \in \ecsg_{(3)} (\forall k \in \nats_{\vert \ecsg_{(3)} \vert}) $
				\vspace*{0.6em}
				\item Send $ (\enc{u_1^{(\numparties)}}, ..., \enc{u_{\vert \ecsg_{(3)} \vert}^{(\numparties)}}) $ to party $ \partysymbol{\numparties - 1} $
			\end{myEnumerate}
			\vspace*{0.4em}
			\item Party $ \partysymbol{i} $ for $ i $ from $ \numparties - 1 $ to $ 1: $
			\vspace*{0.4em}
			\begin{myEnumerate}
				\item Set 
				$ \enc{u_1^{(i)}} := \blind{\enc{u_k^{(i + 1)}} \crossh p^{(i)}_{\ndonorpatient{i}(\ecg_k)}} $
				\\
				with $ \ecg_k \in \ecsg_{(3)} (\forall k \in \nats_{\vert \ecsg_{(3)} \vert}) $
				\vspace*{0.6em}
				\item If $ i \neq 1: $ Send $ (\enc{u_1^{(i)}}, ..., \enc{u_{\vert \ecsg_{(3)} \vert}^{(i)}}) $ to party $ \partysymbol{i - 1} $
			\end{myEnumerate}
			
			\item Party $ \partysymbol{1}: $ \\
			\centering Broadcast $ \enc{L_2} := (\enc{u_1}, ..., \enc{u_{\vert \ecsg_{(3)} \vert}}) := (\enc{u_1^{(1)}}, ..., \enc{u_{\vert \ecsg_{(3)} \vert}^{(1)}}) $
		\end{myEnumerate}
		
		\vspace*{0.4em}	
		\item \underline{Selection Phase}
		\vspace*{0.3em}
		\begin{myEnumerate}
			\item All parties jointly compute $ ((c_1^*, c_2^*)) \leftarrow \crscgate((\enc{L_1}, \enc{L_2})) $
			\vspace*{0.4em}
			\item If $ c_1^* = c_2^* = \bot: $
			\vspace*{0.2em}
			\begin{myEnumerate}
				\item Skip Steps $ 6-7 $
				
				\item All parties output $ 0 $
			\end{myEnumerate}
			\vspace*{0.4em}
			\item Else (\ie, $ (c_1^*, c_2^*) = (\enc{l_1^*}, \enc{l_2^*}) $):
			
			\begin{myEnumerate}
				\item All parties jointly compute $ l_2^* = \dece(\enc{l_2^*}) $
			\end{myEnumerate} 
		\end{myEnumerate}
		
		\vspace*{0.4em}	
		\item \underline{Reverse Mapping Phase}
		\vspace*{0.3em}
		\begin{myEnumerate}
			\item Each party $ \partysymbol{i} \ (i \in \parties): $
			
			\begin{myEnumerate}
				\vspace*{0.4em}
				\item For each $ p_{\ndonorpatient{i}}^{(i)} \in S^{(i)} $ with $ \ndonorpatient{i} \in \ndonorpatientset{i}(\ecsg_{(3)}); $
				\vspace*{0.2em}
				\begin{myEnumerate}
					\item If $ p_{\ndonorpatient{i}}^{(i)} $ divides $ l_2^* $, set $ \ndonorpatientstar{i} := \ndonorpatient{i} $ and go to Step $ 7 $
				\end{myEnumerate}
			\end{myEnumerate}
		\end{myEnumerate}			
		
		\vspace*{0.4em}
		\item \underline{Output Phase}
		\vspace*{0.3em}
		\begin{myEnumerate}
			\item Party $ \partysymbol{i} $ outputs $ \ndonorpatientstar{i} $
		\end{myEnumerate} 
	\end{myEnumerate}
}

%\vspace*{0.3cm}
\end{protocol}

\begin{enumerate}
\item \textit{Construction Phase:}
The parties jointly construct the encrypted adjacency matrix~$ \enc{A} $ encoding the edges of the compatibility graph induced by the parties' input quotes. 
To this end, the parties execute gate~$ \compgate $ for the donor of each party~$ \partysymbol{i} $ ($ \forall i \in \parties $) and the patient of each party~$ \partysymbol{j} $ ($ \forall j \in \parties $) with $ i \neq j $. The corresponding entry $ \enc{a[i,j]} $ of the encrypted adjacency matrix $ \enc{A} $ is determined as the result of the gate execution. In particular, at the end of the construction phase each entry $ \enc{a[i,j]} $ contains an encrypted bit indicating whether the edge $ (i,j) $ is present in the compatibility graph or not. Thus, if $ a[i,j] = 0 $, a transplant between party~$ \partysymbol{i} $'s donor and party~$ \partysymbol{j} $'s patient is not possible.

\item \textit{Evaluation Phase:}
The previously computed adjacency matrix~$ \enc{A} $ is used together with the publicly known set of exchange constellation graphs~$ \ecsg_{(3)} $ (containing exchange cycles up to size $ \cyclesize = 3 $) to determine which of the exchange constellation graphs~$ \ecg_k $ $ (\forall k \in \vert \ecsg_{(3)} \vert) $ are potential exchange constellation graphs. To this end, for all edges of a specific exchange constellation graph~$ \ecg_k $ the corresponding entries of $ \enc{A} $ are multiplied together. The result equals~$ 1 $ if all edges that are present in $ \ecg_k $ are also present in the compatibility graph~$ \cg $ indicating that $ \ecg_k $ is a subgraph of $ \cg $ and thus that $ \ecg_k $ is executable based on the parties' input quotes. Afterwards, each party holds an encrypted binary vector $ \enc{L} $ indicating for each $ \ecg_k $ whether it is executable or not.

\item \textit{Prioritization Phase:}
Since the welfare of an exchange constellation graph in our protocol just corresponds to the number of edges in the graph and since the set of exchange constellation graphs is publicly known, the parties' welfare $ \welfare(\ecg_k) $ of each exchange constellation graph~$ \ecg_k $ ($ \forall k \in \ecsg $) is also publicly known. To obliviously add the welfare of each executable exchange constellation graph to the vector $ L $, the parties can simply multiply the publicly known welfare locally to each entry of the encrypted vector $ \enc{L} $. The resulting encrypted vector is denoted by $ \enc{L_1} $ where an entry $ \enc{L_1[k]} $ is the encrypted welfare of the exchange constellation graph~$ \ecg_k $ if the exchange constellation graph is executable and a fresh encryption of $ 0 $, otherwise.

\item \textit{Mapping Phase:}
Each party $ \partysymbol{i} $ ($ \forall i \in \parties $) is in possession of an interval $ \interval{i} $ of positive integers containing $ \vert \ndonorpatientset{i}(\ecsg_{(3)}) \vert $ distinct prime numbers such that for all $ i, j \in \primeset $ with $ i \neq j $ the intersection of the corresponding intervals is empty, i.e., $ \interval{i} \cap \interval{j} = \emptyset $. Each party $ \partysymbol{i} $ assigns a unique prime number $ \primeNumber^{(i)}_{\ndonorpatient{i}} $ chosen uniformly at random from the set $ \primeset_{\interval{i}} $ of all prime numbers in the interval $ \interval{i} $ to each element of $ \ndonorpatientset{i}(\ecsg_{(3)}) $, i.e., to each possible neighborhood of party~$ \partysymbol{i} $. We denote the set of these prime numbers chosen by a party $ \partysymbol{i} $ by $ S^{(i)} $. The idea is that the parties compute a prime number product for each exchange constellation graph~$ \ecg_k \in \ecsg_{(3)} $ such that each party's contributed prime number encodes the party's neighborhood in that exchange constellation graph. To this end, all parties participate in the computation of the encrypted product $ \enc{u_k} $ with $ k \in \ecsg_{(3)} $ of prime numbers for each exchange constellation graph $ \ecg \in \ecsg_{(3)} $ such that each party supplies one factor $ \primeNumber^{(i)}_{\ndonorpatient{i}} $. The resulting vector of encrypted prime number products $ \enc{L_2} := (\enc{u_1}, ..., \enc{u_{\vert \ecsg_{(3)} \vert}}) $ indicates the neighborhood constellation, i.e., the suggested exchange partners of each party for each exchange constellation graph~$ \ecg_k $. At the end of the mapping phase, $ \enc{L_2} $ is broadcast.

\begin{table*}[!t]
	\centering
\begin{tabular}{L{0.12\linewidth} C{0.085\linewidth} C{0.085\linewidth} C{0.085\linewidth} C{0.085\linewidth} C{0.085\linewidth} C{0.085\linewidth} C{0.085\linewidth} C{0.085\linewidth}}
	Parties					& 2				& 3			& 4 		& 5 		& 6 		& 7			& 8			& 9				\\ \midrule
	Runtime					& $ 14 $s		& $ 24 $s	& $ 44 $s	& $ 2 $m	& $ 6 $m	& $ 26 $m	& $ 2 $h	& $ 13 $h		\\	
	Traffic					& $ 400 $kB		& $ 1 $MB	& $ 3 $MB	& $ 10 $MB	& $ 40 $MB	& $ 200 $MB	& $ 1 $GB	& $ 5.5 $GB		\\					
	$ \vert \ecsg_{(3)} \vert $	& $ 1 $			& $ 5 $		& $ 17 $	& $ 85 $	& $ 275 $	& $ 1211 $	& $ 5915 $	& $ 31067 $		\\		
\end{tabular}
	\caption{Runtimes, network traffic, and size of the exchange constellation graph set~$ \ecsg_{(3)} $ of protocol~$ \aecselprot $ with a maximum cycle size~$ \cyclesize = 3 $.}
	\label{tab:runtimes}
\end{table*}

\item \textit{Selection Phase:}
Using the two previously determined vectors $ \enc{L_1} $ and $ \enc{L_2} $ as input, the parties jointly execute the conditional random selection gate $ \crscgate $ in order to choose the suggested exchange constellation graph~$ \aecg $ uniformly at random from all potential exchange constellation graphs that maximize the welfare, i.e., that allow the maximum number of patients to obtain a kidney transplant. In case there is no such potential exchange constellation graph, i.e., if $ \pecsg = \emptyset $, the parties learn that no suitable exchange constellation graph could be found. Otherwise, the gate returns $ (\enc{l^*_1}, \enc{l^*_2}) \in (\enc{L_1}, \enc{L_2}) $. Afterwards, all parties jointly decrypt $ \enc{l^*_2} $ and, thereby, learn the product of prime numbers encoding the chosen suggested exchange constellation graph~$ \aecg $.

\item \textit{Reverse Mapping Phase:}
Each party $ \partysymbol{i} $ derives its individual exchange partners from the prime number product $ l^*_2 $ by checking which of its prime numbers divides $ l^*_2 $. The neighborhood constellation $ \ndonorpatientstar{i} $ that was mapped to this prime number indicates the suggested exchange partners of party~$ \partysymbol{i} $.

\item \textit{Output Phase:}
Each party $ \partysymbol{i} $ outputs $ \out_i = \ndonorpatientstar{i} $.
\end{enumerate}

The only major change we apply to the privacy-preserving bartering protocol from~\cite{WuellerSemiHonestBartering2017} is the computation of the adjacency matrix in that for the KEP this computation is based on the newly introduced gate~$ \compgate $ for compatibility check. As the correctness and security of this phase relies on the correctness and security of the compatibility check gate $ \compgate $, it is sufficient to prove these properties for gate $ \compgate $ as shown in Section~\ref{sub:comp_prot}. The only other change we apply to the protocol from~\cite{WuellerSemiHonestBartering2017} is that we skip the negotiation phase where the parties determine the quantity of the commodities to barter as for kidney exchange no commodities are necessary. Since the rest of the protocol remains unchanged, we refer to the security and correctness proofs for the bartering protocol provided in~\cite{WuellerSemiHonestBartering2017}.

Similarly, for the complexity analysis we refer to \cite{WuellerSemiHonestBartering2017} except for the construction phase which exhibits a communication complexity~of $ \bigo{\numparties^2 \cdot \bigo{\compgate}} = \bigo{\numparties^3 \cdot \secparam} $ and a round complexity of $ \bigo{\numparties^2 \cdot \bigo{\compgate}} = \bigo{\numparties^3} $. Thereby, we obtain the overall communication complexity $ \bigo{\numparties^3 \cdot \secparam + \numparties^2 \cdot \secparam \cdot \vert \ecsg_{(3)} \vert} $ and round complexity $ \bigo{\numparties^3 + \numparties \cdot \vert \ecsg_{(3)} \vert} $ for the protocol~$ \aecselprot $.

\section{Evaluation}\label{sec:evaluation}

We have implemented our protocol~$ \aecselprot $ (cf.~Protocol~\ref{prot:aecsel}) on top of
the SMC-MuSe framework~\cite{SMCMuseFramework2012} which is 
an SMPC framework that already implements the threshold Paillier cryptosystem~(cf.~Section~\ref{sub:pre_homEnc}) and all gates used in our protocol except for our new gate~$ \compgate $~(cf.~Section~\ref{sub:comp_prot}).

We use a client-server infrastructure where the clients run the actual SMPC protocol and the server only forwards messages between clients and provides them with the keying material of the threshold Paillier cryptosystem.

For the evaluation of our protocol, we have set up a cluster of desktop machines running Ubuntu 16.04 which are connected by a local area network. Each machine is equipped with an Intel Xeon 5400 series CPU and 4GB RAM. One machine runs the server software and each of the other machines runs exactly one client such that each party runs on its own machine. 

The relevant parameters that influence the performance of our protocol are the number of participating parties $ \numparties $ and the key length of the threshold Paillier cryptosystem. We consider numbers of parties from 2 to 9 and a key length of 2,048 bit. We measure the performance of our protocol based on the runtime and the overall network traffic which corresponds to the accumulated incoming traffic of all parties. Each protocol run is repeated 10 times and the results are averaged over the 10 repetitions.

Table~\ref{tab:runtimes} shows runtimes, network traffic, and the size of the exchange constellation graph set~$ \ecsg_{(3)} $ for different numbers of parties. We observe that for small numbers of parties, the protocol exhibits a low runtime, e.g., for up to 5 parties the protocol completes within two minutes. However, we also see that for larger numbers of parties the runtime increases rapidly. For example, for 9 parties the protocol's runtime already amounts to 13 hours. A similar behavior can be observed w.r.t.\ the network traffic. While for 5 parties the network traffic is still only 10 MByte, for 9 parties the traffic already amounts to 5.5 GByte. 

The exponential increase of runtime and network traffic can be attributed to the exponential increase of the set of exchange constellation graphs~$ \ecsg_{(3)} $ which increases with the number of parties. We observe that the size of $ \ecsg_{(3)} $ is small for small numbers of parties, e.g., for 5 parties there are only 85 exchange constellation graphs with cycles up to size $ \cyclesize = 3 $. However, for larger numbers of parties, the size of the set increases drastically. For example, for 9 parties the set already includes 31,067 graphs.

\begin{figure}[h]
	\centering
\begin{tikzpicture}[scale=0.73]
\begin{axis}[
	 x tick label style={/pgf/number format/1000 sep=},
	 enlargelimits= 0,
	 height = 5cm,
	 width = 9cm,
     legend style={at={(1.04,0.7)},anchor=north west},
	 ybar interval=0.7,
	 xlabel= Number of parties,
	 ylabel= {Runtime [\%]},
     xmin= 2,
     xtick= {2,3,4,5,6,7,8,9, 10, 11},
     ytick= {0.2, 0.4, 0.6, 0.8, 1.0},
     xmax= 10,
     ymin= 0,
     ymax= 1,
]
\addplot[color=smartred, fill = smartred, area legend] coordinates {(2, 0.94)(3, 0.83)(4, 0.65)(5, 0.38)(6, 0.15)(7, 0.04)(8, 0.01)(9, 0.00)(10, 0)};

\addplot[color=smartblue, fill = smartblue, area legend] coordinates {(2, 0.03)(3, 0.05)(4, 0.08)(5, 0.11)(6, 0.12)(7, 0.12)(8, 0.12)(9, 0.12)(10, 0)};

\addplot[color=smartgreen, fill = smartgreen, area legend] coordinates {(2, 0.01)(3, 0.09)(4, 0.24)(5, 0.47)(6, 0.68)(7, 0.79)(8, 0.83)(9, 0.84)(10, 0)};

\legend{Construction, Evaluation, Selection}
\end{axis}
\end{tikzpicture}
\vspace{-0.5cm}
	\caption{Split of the runtime for the most dominant phases (construction, evaluation, selection) of the protocol~$ \aecselprot $ consumes for increasing numbers of parties. The construction phase shows the influence of our compatibility check gate~$ \compgate $ on the protocol~$ \aecselprot $.}
	\label{fig:phase_comp}
\end{figure}
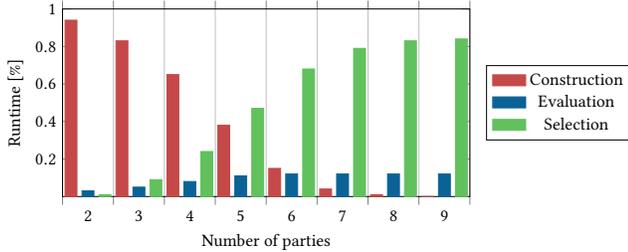

The runtime consumed by each phase of the protocol also indicates that the size of the exchange constellation graph set $ \ecsg_{(3)} $ directly correlates with the performance of the protocol $ \aecselprot $. Figure~\ref{fig:phase_comp} shows the percentage of the overall runtime consumed by each of the protocol phases for different numbers of parties. The bar chart only includes the three phases construction, evaluation, and selection as the other phases only consume negligible amounts of time. We observe that with an increasing number of parties, the impact of those phases whose performance increases with the size of the exchange constellation graph set increases also. While the size of the adjacency matrix that is computed during the construction phase only increases quadratically with the number of participating parties, the evaluation phase requires a call to the gate~$ \ufimultgate $ for each exchange constellation graph and thus its performance increases with the size of the set of exchange constellation graphs. Also, the performance of the selection phase depends on the size of~$ \ecsg_{(3)} $ as the size of the input vectors $ L_1 $ and $ L_2 $ to the gate~$ \crscgate $ for conditional random selection is determined by the number of exchange constellation graphs in $ \ecsg_{(3)} $. The execution of the selection phase clearly dominates the runtime of the other phases for larger numbers of parties. Furthermore, Figure~\ref{fig:phase_comp} shows that the construction phase only has a small impact on the overall protocol performance for increasing numbers of parties. Recalling that the construction phase only consists of calls to our novel gate~$ \compgate $ for compatibility check in kidney exchange, we can deduce that the impact of our gate~$ \compgate $ on the overall performance of the protocol~$ \aecselprot $ is also small for increasing numbers of parties.

This leads to the conclusion that the performance bottleneck of our kidney exchange protocol on the one hand is the implementation of conditional random selection and on the other hand is due to the large size of the exchange constellation graph set~$ \ecsg_{(3)} $. Thus, a more efficient implementation of conditional random selection would directly lead to more efficient runtimes of our protocol also for larger numbers of parties. Despite the long runtime of our protocol for larger numbers of parties, the low runtimes for small numbers of parties allow for the application of our protocol in a dynamic kidney exchange system where the protocol is repeatedly executed among small sets of parties. 

Furthermore, it is important to note that for the application of kidney exchange the protocol runtime is not critical as the KEP does not have to be solved in real time. Instead, in practice living donor programs in many countries (re-)evaluate incompatible patient-donor pairs for exchange possibilities rather infrequently, typically only once every few months~\cite{Biro_EuropeanKE_2019}.

\section{Conclusion and Future Work}\label{sec:conclusion}

In this paper, we have demonstrated that it is possible to devise a privacy-preserving protocol for the KEP that exhibits very efficient runtimes for a fixed small number of parties. Also, since non-privacy-preserving algorithms that are currently used to solve the KEP do not run in real time but instead are typically only executed once every few months, it is not an impediment in practice that the runtime of our privacy-preserving protocol increases considerably with a growing number of parties.

For future work we plan to pursue two main directions. First, we aim to devise an overarching kidney exchange system that will use our privacy-preserving protocol for the KEP at its core. From a system perspective it is necessary to handle arriving and departing parties as well as parties participating in more than one protocol execution. This poses the challenge of defining a suitable privacy notion outside of the traditional SMPC context. Second, more recent non-privacy-preserving kidney exchange algorithms also include so-called altruistic donors and provide for some means of redundancy to mitigate the effects of last-minute failures (e.g., due to the drop-out of a donor). We plan to investigate whether it is possible to implement such features in a privacy-preserving fashion.

\begin{acks}
	This work was funded by the Deutsche Forschungsgemeinschaft (DFG, German Resarch Foundation) - project number (419340256) and NSF grant CCF-1646999. Any opinion, findings, and conclusions or recommendations expressed in this material are those of the author(s) and do not necessarily reflect the views of the National Science Foundation.
	
	Computations for our SMPC protocols were performed with computing resources granted by RWTH Aachen University under Project RWTH0438.
\end{acks}

\bibliographystyle{ACM-Reference-Format}
\bibliography{references}

\appendix

\end{document}